**Kinesin-driven de-mixing of cytoskeleton composites drives emergent mechanical properties**

Janet Sheung[1], Christopher Gunter[2], Katarina Matic[3], Mehrzad Sasanpour[3], Jennifer L. Ross[4], Parag Katira[2], Megan T. Valentine[5], Rae M. Robertson-Anderson[3*]

[1]Department of Natural Sciences, Scripps and Pitzer Colleges, Claremont, CA, United States; W. M. Keck Science Department, Claremont Mc College, Claremont, CA 91711, USA
[2]Department of Mechanical Engineering, San Diego State University, San Diego, CA 92182, USA
[3]Department of Physics and Biophysics, University of San Diego, San Diego, CA 92110, USA
[4]Department of Physics, Syracuse University, Syracuse, NY 13244, USA
[5]Department of Mechanical Engineering, University of California, Santa Barbara, Santa Barbara CA 93106, USA

*Corresponding Author: randerson@sandiego.edu

**ABSTRACT**

The cytoskeleton is an active composite network of protein filaments, such as microtubules and actin filaments, and motors, such as kinesin, that dictate the mechanical properties and processes in eukaryotic cells. To achieve the diverse mechanics necessary for these functions, the cytoskeleton actively restructures itself, often by enzymatic motors generating athermal forces to pull on filaments. This restructuring leads to spatiotemporally heterogeneous force responses that are critical to cellular multifunctionality but render mechanical characterization very challenging. In this work, we couple optical tweezers microrheology and fluorescence microscopy with simulations and mathematical modeling to robustly characterize the mechanical response of in vitro composite networks of co-entangled microtubules and actin filaments undergoing active restructuring by kinesin motors that crosslink and exert forces between microtubules. We discover that active composites exhibit a rich ensemble of force response behaviors that can be classified as elastic, yielding, and stiffening, with the propensity and properties of each tuned by the kinesin concentration and local strain rate. Our results reveal emergent mechanical stiffness and resistance at intermediate kinesin concentrations with higher concentrations exhibiting lower stiffness and more viscous dissipation. Structural analysis of simulated and experimental composites reveals that actin and microtubules transition from a well-mixed interpenetrated double-network to a de-mixed state of microtubule-rich aggregates surrounded by actin phases that are relatively undisturbed. It is this de-mixing that leads to the emergent force resistance of the composite, offering an alternate route that composites can leverage to achieve enhanced stiffness through coupling of structure and mechanics.

**INTRODUCTION**

The cytoskeleton is an active composite network of filamentous proteins and their associated binding proteins, including energy-transducing molecular motors that pull and walk along filaments[1,2]. A primary role of the cytoskeleton is to provide mechanical integrity to cells while also allowing them to stiffen, soften, change shape, and generate forces, often in response to local stimuli[3,4]. These diverse mechanical responses are often spatially heterogeneous and can range from nanoscopic to cell-spanning scales. Moreover, the nature of the response is intricately linked to the time-evolving structures and interactions of the different networks of e.g., semiflexible actin and microtubules[5].

This complex active and composite nature of the cytoskeleton has rendered it a foundational model system for probing questions in active matter physics and addressing design challenges in living materials[3,6]. In vitro cytoskeleton-based active matter systems[7–15] typically include myosin II minifilaments[16] and/or crosslinked clusters of kinesin dimers[17], which are enzymatically-active motor proteins that harness the energy of ATP hydrolysis to bind to and pull on actin filaments and/or microtubules, respectively. Actomyosin networks have been shown to undergo bulk contraction, local contraction into foci or asters, or disordered flow depending on the concentrations of the myosin, actin and crosslinkers[12,13,18]. Kinesin clusters acting on bundles of microtubules have also shown varied behaviors, ranging from the formation of locally condensed asters[19–23] to space-spanning networks capable of extensile restructuring that results in nematic flow and organization reminiscent of liquid crystals[8,24,25].

More recently, in vitro active cytoskeletal composites that include both actin and microtubules have been engineered and examined[26–33], often revealing emergent behavior and improved material properties, such as organized dynamics[26], tunable miscibility[30,33], structural memory[31], and enhanced elasticity[28]. Early work explored passive composite networks lacking molecular motors. In this simplified condition, biotin-streptavidin crosslinkers induce passive and effectively permanent crosslinking of either the actin or microtubule components[34,35]. Within such composites, microtubule crosslinking is essential to eliciting elastic responses to localized strains, whereas actin-crosslinked composites exhibit yielding behavior similar to that of purely entangled composites[34].

When the concentration of actin crosslinkers was varied, an emergent elasticity was revealed at intermediate crosslinker:actin ratios $R \simeq 0.2$, which decreased to values comparable to those of entangled composites as this ratio was increased to $R \simeq 0.8$ [35]. This counter-intuitive behavior is observed only in composites, and is driven by crosslinker-mediated network coarsening and bundling that simultaneously increases the thickness (and thus stiffness) of network fibers and as well as network mesh size. The delicate interplay of actin network microstructure and fiber rigidity leads to the development of an optimal crosslinker ratio in which the network has developed sufficient rigidity to maximize elastic response, but the mesh size remains small enough to suppress the diffusive mobility of microtubules entrapped within the actin network[35]. Notably, for actin-only networks, increasing crosslinker ratios monotonically increases the network stiffness[36–38]. However, within the composites, the increased microtubule mobility enables new pathways of stress relaxation, which dominate the mechanical response and soften and fluidize the composite.

When enzymatically-active motors are included, even more dramatic structural changes are observed. Embedded motors act both as transient crosslinkers and force-generating elements, leading to dynamic responses that span spatial and temporal scales. Co-entanglement of microtubule filaments with myosin-driven actin produces active composite networks in which both actin and microtubules ballistically contract at speeds that can be tuned by the concentrations of actin and myosin[26,27]. Such networks display controlled motion, enhanced elasticity, and sustained structural integrity as compared to single filament networks[28]. Replacing myosin with kinesin as the active agent in similar co-entangled composites results in a much higher degree of variability in dynamics and structure, with speeds that vary by 2 orders of magnitude, phases of acceleration and deceleration, and enhanced restructuring and de-mixing of actin and microtubules, depending on the composite formulation and time after motor activation[33]. Typically, network restructuring occurs over a finite timespan that is determined by the motor-driven compaction of filaments into poorly-connected, kinetically trapped network structures reminiscent of asters or disordered aggregates, with kinesin-driven composites displaying shorter active lifetimes as compared to those driven by myosin or both myosin and kinesin.

Similar phenomena have been observed in composite networks comprising kinesin clusters and microtubules that are bundled by osmotic crowders acting as depletants and by microtubule binding proteins that promote antiparallel bundling[30]. Adding low concentrations of actin to such networks produces fluid-like extensile dynamics, similar to those of kinesin-driven microtubule networks lacking actin. However, when the actin concentration is increased, rich dynamic structural transitions are observed, leading to the formation of onion-like asters of layered actin and microtubules or bulk contractility. Further increases in actin concentration promote de-mixing of actin and microtubules with asters and contractile regions becoming increasingly microtubule-rich[30].

Motor-driven structural transitions that form disconnected, filament-dense structures interspersed within a dilute fluid phase can undermine network percolation. From a material design perspective, this phase segregation risks fluidizing the network on mesoscopic scales, thereby compromising the ability of active composites to transmit forces over large distances or to sustain significant external stresses. Despite the importance of the mechanical response of these systems to their role in cellular processes and to materials-based applications, the majority of active composite studies have focused on the evolving structure and dynamics of the materials without regard to mechanical properties.

While there are a number of approaches to analyzing network structural rearrangements via fluorescence microscopy, the measurement of the time-evolving local mechanical properties within the dynamically-restructuring network remains technically challenging. The emergent heterogeneity arising due to motor-driven aggregation or segregation of filaments demands the precise application of forces to determine the local mechanical responses,[28] as well as relatively large numbers of measurements in order to develop an understanding of average responses and the range of variations at each condition. Thus, although a key feature of motor-driven active cytoskeletal composites is their ability to flow, coarsen and reconfigure due to internal motor-generated forces, little is known about the interplay between material mechanics and filament motion during the emergence of these new structural phases.

To begin to establish the foundational role of motor-driven restructuring in network mechanics, we designed co-entangled composites of actin and microtubules formulated to support robust connectivity, and subjected them to active stresses and restructuring by adding varying concentrations of kinesin motors. Using a comprehensive platform comprising an optical tweezers microrheometer (OTM) capable of applying large-scale strains at specified locations within the heterogeneous sample, fluorescence microscopy to assess structural rearrangements, simulations based on lattice-based advection-diffusion models, and mathematical modeling of mechanical responses, our results reveal how kinesin motors act on composites of actin and microtubules to sculpt the mechanical and structural properties across spatiotemporal scales. We identify the presence of kinesin-driven demixing via clustering, which in turn leads to emergent complexity in mechanical response and formulation-dependent heterogeneity that can be captured both in vitro and in silico. These results demonstrate the importance of hierarchical structural heterogeneity to provide new avenues for enhanced stiffness and relaxation only possible in composite designs.

**RESULTS & DISCUSSION**

**Kinesin motors drive de-mixing via clustering of microtubules**

As a first step, we assess the complex structural properties of active composites composed of actin, microtubules, and kinesin. We judiciously chose a ratio of actin to microtubules (45:55 molar ratio of actin to tubulin dimers) that allows for active restructuring and force-generation without the large scale flow or network rupturing that has been previously reported[30,33], and examined the effect of varying concentrations of kinesin, $c_k$, on the network restructuring (**Fig 1A**). In control networks lacking kinesin, we observed, using high-resolution two-color confocal microscopy, uniform mixing of actin and microtubules, to form a homogeneous, space-spanning composite of interpenetrating networks of actin and microtubules (**Fig 1A, left**). Upon addition of kinesin, we observed the formation of microtubule-rich phases that appeared to generally increase in size, density and number with increasing kinesin concentration. This kinesin-driven de-mixing of microtubules from actin was robustly observed across all samples; and, upon addition of 640 nM kinesin, the highest concentration investigated here, nearly all of the microtubules condense into aster-like aggregates surrounded by actin-rich zones (**Fig. 1A, right**).

To better understand the molecular drivers of this behavior, we developed a two-dimensional lattice-based advection-diffusion model of filament dynamics. Within the model, filament motion arises from kinesin-generated active forces that can either pull or push microtubules, as well as frictional forces that occur when passive motors act as crosslinkers between microtubules[33]. In the control case without kinesin, the simulations rendered a uniform, well-mixed composite of interpenetrating networks of microtubules and actin, consistent with our experimental observations (**Fig 1B, left**). Similarly, upon addition of kinesin motors, the composites restructure and de-mix, with increasing segregation observed for higher kinesin concentrations. Moreover, it is clear from both experiment and simulations that the kinesin-driven motions that cluster the microtubules do not significantly restructure the actin. Rather, a two-phase material is formed, with microtubule-dense regions forming distinct, well-separated aggregates within a more uniform actin-rich background (**Fig 1B, right**).

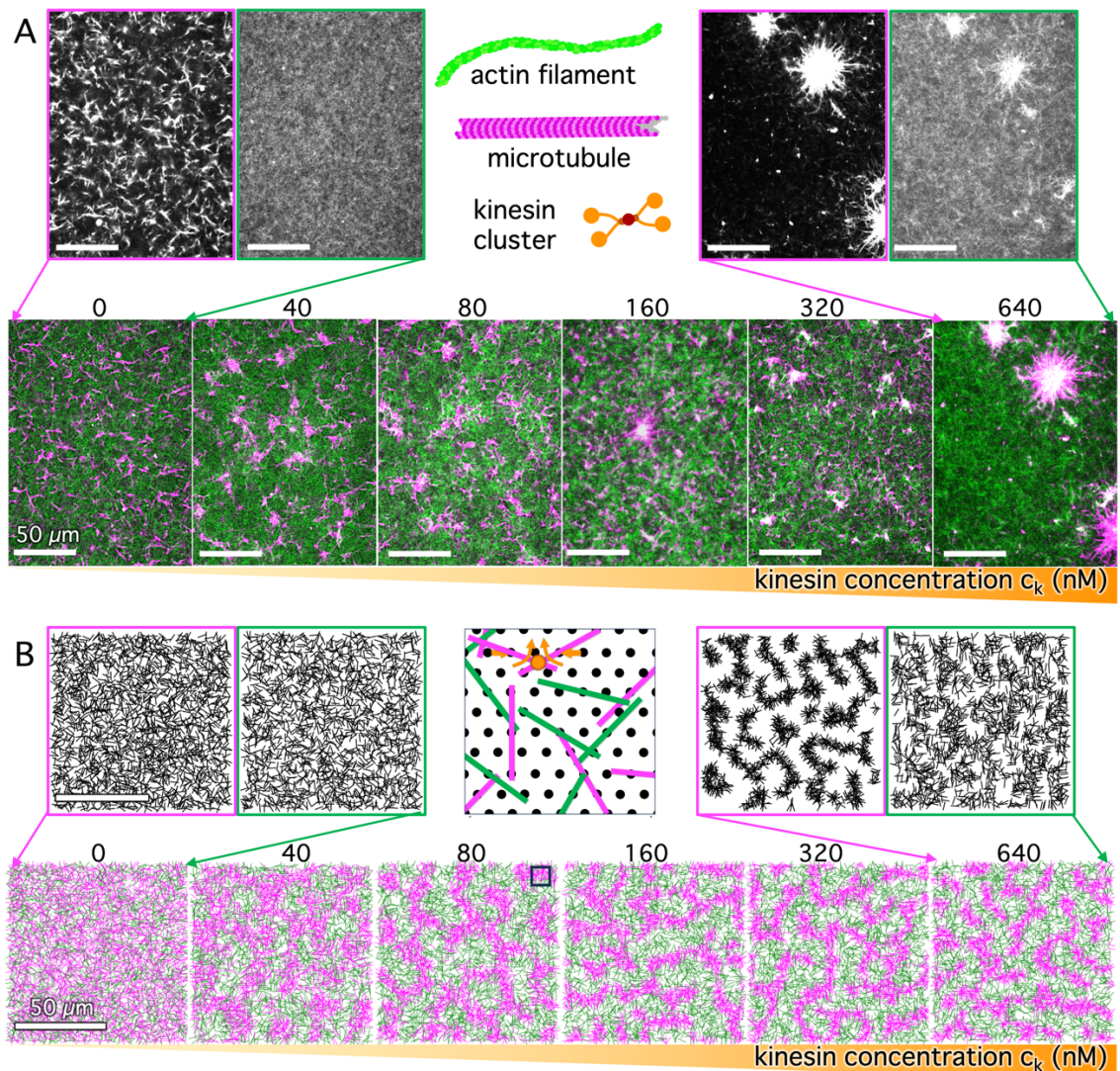

**Figure 1. Kinesin motors drive de-mixing of co-entangled actin and microtubules**. (A) Two-color fluorescence confocal microscopy images of composites of microtubules (magenta) and actin filaments (green) in the presence of varying concentrations of kinesin motors, listed in nM above each composite image. Greyscale images show separate channels for microtubules (magenta borders) and actin (green borders) for $c_k = 0$ nM (left) and $c_k = 640$ nM (right). All scale bars denote 50 μm. Schematics show composite components (not to scale). (B) Snapshots from simulations of 2D kinesin-driven composites of actin and microtubules with the same effective kinesin concentrations as in experiments. Colors and labels are the same as in A. Center schematic is a zoom-in of the simulated composite with black circles denoting lattice points that can be occupied (or not) by a microtubule (magenta) or actin filament (green). The orange circle and arrows denote, respectively, a kinesin motor and filament rotation (curved arrows) and contraction (straight arrows) that it can impart on microtubules. The scale of the schematic is indicated by the black box in the upper right corner of the $c_k = 80$ nM snapshot.

The simulations also provide the opportunity to quantitatively compare the simulated network structures before and after restructuring through calculation of the filament pair distribution function $g_{ij}(r,T)$ where the subscripts $i$ and $j$ represent the filament type, either actin (A) or microtubules (M), and which gives the probability of finding a filament (actin or microtubule) a radial distance $r$ from any other filament. To assess the dynamic structural changes that occur within a single filament network during the simulation, we report the difference between these quantities for the initial ($T = 0$) state and final ($T = T_F$) states: $\Delta g_{ii}(r) = g_{ii}(r,T_F) - g_{ii}(r,0)$. For static, steady state networks, we expect $\Delta g_{ii}(r) = 0$ for all $r$, as we see in **Fig 2A,B** for both the actin and microtubule networks within the composites lacking kinesin ($c_k = 0$). This result also validates that our initial simulation conditions represent homogeneous, well-mixed networks that remain well-mixed in the absence of motor activity. When comparing the distribution of microtubules to other microtubules, or actin filaments to other actin filaments, we found positive values of $\Delta g_{MM}(r)$ and $\Delta g_{AA}(r)$ for small filament separation distances $r$, indicating attractive interactions that drive clustering on those length scales; and we observed an increased clustering propensity with increased kinesin concentration, estimated by the value of $\Delta g_{ii}(r_0)$, where $r_0 = 1.25$ μm is the smallest radial distance between lattice points in the simulation (**Fig. 2D**). As the separation distance increases, $\Delta g_{ii}(r)$ curves for all $c_k > 0$ composites decay to zero then continue to decrease, reaching local negative-valued minima before asymptoting back to zero. This behavior indicates a depletion of filaments on intermediate length scales, which we interpret as an indication of clustering. While both filament types display these general features, the microtubule network, upon which the kinesin motors directly act, exhibited much stronger clustering effects compared to actin (**Fig 2A,B,D**).

By contrast, when evaluating the co-distribution of actin filaments with respect to the microtubule network, we find negative values of $\Delta g_{MA}(r)$ at even the smallest filament separation distances (**Fig 2C**), indicating an exclusion of the unlike filament type. This anti-correlation demonstrates that actin is displaced from the microtubule-rich domains that form from the kinesin-driven contraction of microtubules, and is consistent with de-mixing. The strength of this effect can be approximated by $\Delta g_{MA}(r_0)$, which is found to monotonically decrease with increasing kinesin concentration (**Fig 2D**). The magnitude of this exclusion effect $|\Delta g_{MA}(r_0)|$ is intermediate between the values observed for clustering of the microtubule $\Delta g_{MM}(r_0)$ and actin $\Delta g_{AA}(r_0)$ structures. The length scales over which this phase separation was observed can be approximated by the radial distance at which $\Delta g_{ij}(r) = 0$, which we denote as $l_0$, as well as the distance at which $\Delta g_{ij}(r)$ is minimal for like-filament distributions or maximal for microtubule-actin co-distributions, which we denote by $l_{min}$ or $l_{max}$. Specifically, $l_0$ can be considered a measure of cluster size while $l_{min}$ and $l_{max}$ are measures of spacing between clusters. In a system with mass conservation, we expect both quantities to generally track with one another, as we observe in Fig 2E.

As shown in **Fig 2E**, we found similar length scales of de-mixing when comparing all filament types, and, in each case, we observed a monotonic decrease in the observed length scales with increasing kinesin concentrations. Moreover, the range of values (~5 – 15 μm) were generally consistent with the observed sizes and spacing between clusters in both experiment and simulation (**Fig 1**), and reflect the increased clustering with increasing kinesin concentration.

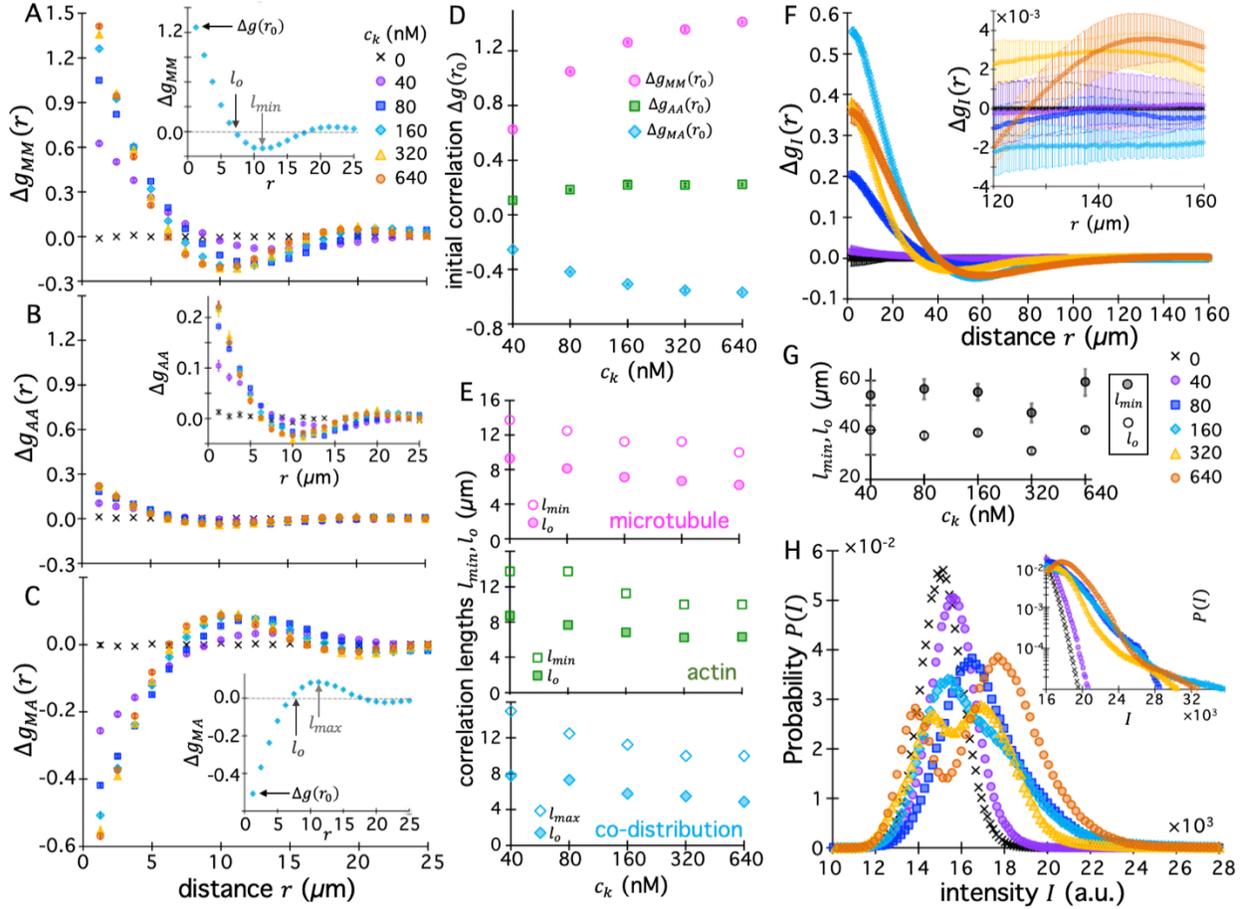

**Figure 2. Filament pair correlations and clustering of microtubules increase with increasing motor concentration over varying lengthscales**. (A-C) Correlation analysis of simulated composites at varying kinesin concentrations (listed in legend in A) show increased correlation between (A) microtubule pairs $\Delta g_{MM}(r)$ and (B) actin pairs $\Delta g_{AA}(r)$ as well as (C) decreased co-distribution of actin and microtubules at short distances $r$ compared to the no kinesin case ($c_k = 0$, black × markers). Insets in A and C depict metrics plotted in D and E. Inset in B is zoom-in of $\Delta g_{AA}(r)$. (D) Initial values of distributions plotted in A-C, $\Delta g_{MM}(r_0)$ (magenta circles), $\Delta g_{AA}(r_0)$ (green squares), and $\Delta g_{MA}(r_0)$ (blue diamonds), show increasing like-filament correlation and decreasing co-distribution as $c_k$ increases. (E) Correlation lengths determined as the radial distances $r$ at which $\Delta g_{MM}(r)$ (top, magenta), $\Delta g_{AA}(r)$ (middle, green) and $\Delta g_{MA}(r)$ (bottom, blue) reach zero ($l_0$, filled symbols) and local extrema ($l_{min}, l_{max}$, open symbols). (F,G) Spatial image autocorrelation analysis of experiment videos of labelled filaments, showing the (F) average autocorrelation difference of pixel intensities $\Delta g_I(r)$ for varying kinesin concentrations (see legend below), and (G) the corresponding correlation lengths $l_0$ (open symbols) and $l_{max}$ (filled grey symbols) versus $c_k$. Inset in F shows zoom-in of $\Delta g_I(r)$ curves at large distances with error bars (which are too small to see in the main plot) denoting standard error. (H) Probability distributions of pixel intensities $P(I)$ for varying kinesin concentrations. Inset shows distributions on a semi-log scale to better visualize the high-intensity tails at high kinesin concentrations ($c_k > 80$ nM).

To more quantitatively compare these structural analysis results from simulations to experimentally observed restructuring, we performed spatial image autocorrelation (SIA) analysis[28,39] on epifluorescence movies of the composites captured in the same samples as the force measurements that we describe below. In SIA the correlation in intensities between two pixels separated by a distance $r$ is examined. In this experiment, both actin and microtubules were labeled with spectrally-indistinguishable fluorescent dyes and were simultaneously imaged such that at each condition, the composite network behavior is observed (see Methods). Similar to pair distribution functions obtained from simulations, the resulting autocorrelation function $g_I(r)$, decays from a maximum value at $r_{0,I} = 0.41$ µm (set by the pixel size) to $g_I = 0$ as $r \to \infty$, passing through a local $g_I < 0$ minimum; and at a given distance larger $g_I(r)$ values are suggestive of increased filament clustering. To better compare the simulated distributions shown in **Fig 2A-C** to experimentally determined data, we subtracted the value of $g_I(r)$ obtained for the composite without kinesin ($c_k = 0$) from $g_I(r)$ for each $c_k > 0$ composite, similar to subtracting the initial time distribution from the final in simulations. As shown in **Fig 2F**, the resulting $\Delta g_I(r)$ curves display similar functional features as the simulated data, with higher kinesin concentrations generally resulting in larger $g_I(r_0)$ values and more pronounced minima. However, a striking distinction is that for experiments, $g_I(r_0)$ displays a non-monotonic dependence on $c_k$, reaching a maximum for $c_k = 160$ nM. This non-monotonicity is reminiscent of similar emergent phenomenon reported for cytoskeleton composites with increasing concentrations of actin crosslinkers[35].

Evaluating the same characteristic distances as in simulations for cluster size and spacing, $l_0$ and $l_{min}$, namely where $\Delta g_I(r)$ reaches zero and a local miminum, we find that both lengthscales show a modest decrease between $c_k = 40$ nM and $c_k = 320$ nM, similar to simulations, but subsequently increases at $c_k = 640$ nM (**Fig 2G**). This increase is indicative of the large aggregates we observe in microscopy images (**Fig 1A**). We note that the correlation lengthscales observed in experimental data (~30 – 60 µm) are generally larger than for simulations, likely due to the larger field-of-view and system size, as well as the added dimension in 3D experiments. Further examining the large lengthscales accessible to experiments, we observed positive $g_I(r)$ values out to the largest analyzed distance ($r_\infty = 160$ µm), for the highest kinesin concentrations ($c_k = 320, 640$ nM), indicative of the presence of largescale clustering in these conditions (**Fig 2F, inset**). By contrast, the $c_k = 160$ nM composite, which displayed the highest short-range correlation, exhibited negative long-range correlation values as $r \to r_\infty$. Together, these data suggest that as the kinesin concentration increases, de-mixing initially causes dense small-scale clustering, as indicated by the peak in $g_I(r_0)$ at intermediate kinesin concentration, followed by large-scale phase separation that maximizes $g_I(r_\infty)$ for higher kinesin concentrations.

To further corroborate the physical picture of de-mixing, we also examined the distribution of pixel intensities of the same videos. Using intensity as a proxy for mass, we evaluated the distribution of pixel intensities to identify increases (higher pixel values) and decreases (lower pixel values) in filament density due to bundling and clustering (**Fig 2H**). We found that as the kinesin concentration was increased from $c_k = 0$ nM to 80 nM, the peaks of the distribution shifted to higher intensity values, indicating bundling; and the distributions became broader, indicating the increasingly heterogeneous distribution of densities. For $c_k \geq 160$ nM, two peaks emerged. The higher intensity peak occurred at an intensity that was slightly larger than that of the single peak for the $c_k < 160$ nM conditions, and this peak shifted to higher intensity values (shifting further right) and larger probabilities (increasing height) as $c_k$ increased. This trend is indicative of an increasing number of bundles that also become denser due to the presence of additional kinesin motors. The second peak, which occurred at lower intensity values than the single peaks observed for $c_k < 160$ nM, likewise shifted to lower intensity values as $c_k$ increased, indicating the emergence of more microtubule-poor zones. Zooming-in to examine the high intensity tails of the distributions, we found that composites with $c_k \geq 160$ nM exhibited pronounced extended tails that were not observed for the lower kinesin concentrations, again indicating the formation of large and dense clusters at the higher concentrations of kinesin (**Fig. 2H, inset**).

Together, these results indicate that kinesin clusters drive contraction and compaction of disordered microtubules into dense, well-separated aggregates. This contraction occurs in the absence of osmotic crowding agents and does not require the presence of non-motor microtubule associated binding proteins to promote bundling. This restructuring causes modest reorganization of the actin network, as the actin filaments are squeezed out by contracting microtubules. However, the actin network remains reasonably well dispersed even at the highest kinesin concentrations.

**De-mixing drives emergent complexity in mechanical response**

To probe how the kinesin-driven restructuring influences the microscale mechanical properties of the composite, we applied localized but large-scale deformations within the heterogeneous material using an optical trapping-based manipulation platform (**Fig 3A,B**)[40–42]. A single beam gradient optical trap was formed by tightly focusing a high-powered IR laser to a diffraction-limited volume within the sample chamber[43]. This allowed the capture and manipulation of embedded colloidal probes (radius $r_p = 2.25$ µm, full details provided in Materials and Methods). The trap stiffness, which was calibrated through independent measurements, was sufficient to stably trap and hold the particle, even as the stage moved at fixed velocity, thereby dragging the particle through the sample. The displacement of the trapped particle from the trap center was simultaneously monitored in real time, and when multiplied by the known trap stiffness, provided an instantaneous readout of the force. From the known values of force and stage position, which are collected as a function of time (**Fig. 3C**), it is possible to construct a relationship between force and stage position. Thus, this instrument acts as a microscale mechanical testing system or microrheometer, which we use to interrogate the response to a strain (i.e., stage displacement) of $s = 20$ µm, which we chose to be significantly larger than the probe size $r_p$ and composite mesh size $\xi \simeq 1.2$ µm (see Methods). We performed measurements on composites with the six kinesin concentrations presented above (**Figs 1,2**), using 3 stage speeds ($v = 6, 12, 24$ µm s$^{-1}$) for each concentration.

Examining the individual force-displacement traces (**Fig 3C,D**), we find that all traces exhibited sharp initial increase in force at the smallest stage displacements as the particle position rapidly shifted within the trap as the stage began to move. The time constant associated with this re-equilibration is given by the ratio of local drag coefficient to the trap stiffness $k_{OT}$ and was typically < 50 ms. Beyond this very initial behavior, we find heterogenous responses, which can be categorized into three broad classes of responses (shown in representative traces in **Fig. 3D**): traces that are fully linear, suggesting elastic behavior ('elastic'), those that show an initial elastic response that softens or yields over time ('yielding'), and those that show an initially soft elastic response that significantly stiffens at large displacements ('stiffening'). The proportion of traces falling into each category varies as a function of kinesin concentration and stage speed (**Fig. 3E**). In general, the responses are diverse and heterogeneous, likely reflecting the structural heterogeneity of the material observed in Figures 1 and 2. The observance of a large fraction of fully linear elastic traces is notable, given the stage stroke of 20 μm, an order of magnitude larger than the size of both the probe and mesh.

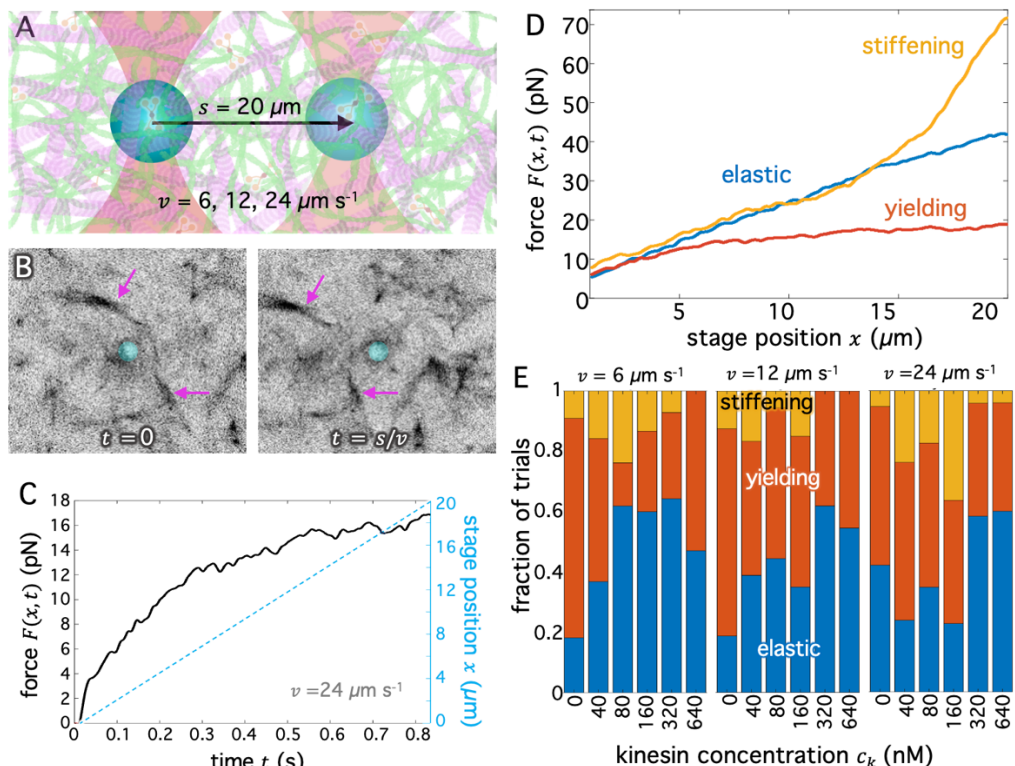

**Figure 3. Optical tweezers microrheology reveals heterogeneous distribution of force responses of kinesin-driven composites to local mesoscopic strains.** (A) Schematic showing a focused infrared laser (red) trapping a probe particle of diameter 4.5 μm (blue) embedded in a kinesin-driven actin-microtubule composite. The sample is deformed locally by the application of a constant-speed strain operating over a of distance $s = 20$ μm at speeds $v = 6, 12, 24$ μm s$^{-1}$, by the action of a piezoelectric stage moving the sample relative to the trap. (B) Inverted greyscale images of labelled fibers in the composite being displaced (magenta arrows) as the stage moves relative to the fixed trap that holds the probe (highlighted in cyan). The images show the time immediately before ($t = 0$, left) and after ($t = \frac{s}{v}$, right) the initial stage sweep; magenta arrows highlight structures visible in both images to demonstrate the relative motion of the trapped particle with respect to the surrounding composite. (C) Example of the force (black) exerted on the probe versus time $t$ in response to the stage (sample) moving through a displacement (cyan, dotted line) of $s = 20$ μm relative to the trap at a speed of $v = 24$ μm s$^{-1}$. (D) Representative examples of force-displacement curves demonstrating the 3 classes of responses that composites exhibit: elastic (blue), yielding (orange), and stiffening (gold). (E) Fraction of trials that exhibited each response class, color coded as in panel D, for all 6 kinesin concentrations ($x$ axis) and all 3 speeds: 6 μm s$^{-1}$ (left), 12 μm s$^{-1}$ (middle), and 24 μm s$^{-1}$ (right). The total number of measurements per condition varied from 11 to 25.

We find that stiffening responses are the least likely across all kinesin concentrations and speeds, and the largest proportion of stiffening traces is found at intermediate kinesin concentrations, perhaps indicating an increased likelihood of microtubule bundling, but without large-scale separation into dense aster-like aggregates, which is most prominent at the highest kinesin concentrations (**Fig. 2F,H**). Consistent with this interpretation, in some cases, the initial slope of the force-displacement curve in the stiffening traces is nearly zero, suggesting that the beads are moving through a very weak or viscous material, with stiffening occurring only when the bead encounters a dense cluster or aggregate of filaments. We note that these results differ from those obtained using a similar method to interrogate the mechanical properties of steady-state actin-microtubule composites lacking motors, which show largely yielding or elastic behavior in the large deformation regime, depending on the concentration of passive crosslinkers[34,44].

To better assess the dependence of mechanical properties on composite formulation, we analyzed the force-displacement traces measured for >20 different particles in different locations and samples for each experimental condition (see Methods). For each combination of kinesin concentration $c_k$ and speed $v$, we averaged together all traces that displayed elastic, yielding, or stiffening characteristics (**Figs 4A**, **S2**). For each of these response types, we observed striking nonmonotonic behavior as the kinesin concentration was increased.

When we examined the subset of elastic traces obtained at $v = 24$ µm s$^{-1}$, we found that the largest value of maximum force and maximum effective stiffness, as qualitatively assessed from the terminal force value $F_{max}$ reached at the end of the strain, occurred at the intermediate value of $c_k = 160$ nM (**Fig. 4A,B**). The lowest values of $F_{max}$ were surprisingly observed at the highest concentration of $c_k = 640$ nM. Similar general trends were observed for the elastic traces obtained at the other speeds (**Fig. 4B**, **Fig S2**).

For yielding traces observed at $v = 24$ µm s$^{-1}$, the measured force typically settled to a plateau value for stage displacement values above ~5 µm (**Fig 4A**), which corresponds to a local strain of approximately 1 if we estimate the local strain by normalizing the stage displacement by the trapped particle diameter[40,45]. The maximum force value initially increased with increasing concentrations of kinesin until the intermediate value of $c_k = 80$ nM was reached, after which the $F_{max}$ showed a slight decline, as shown in **Fig 4B** where we display $F_{max}$ normalized by the $c_k = 0$ nM value. Similar trends were observed for the yielding traces obtained at $v = 12$ µm s$^{-1}$, whereas at $v = 6$ µm s$^{-1}$, $F_{max}$ increased monotonically with increasing concentrations of kinesin (**Fig 4B**). Among the particles exhibiting stiffening at $v = 24$ µm s$^{-1}$, we observed a peak in maximum force at $c_k = 80$ nM, which then decreased to a value lower than that of the initial $c_k = 0$ nM condition at the largest concentration of $c_k = 640$ nM (**Fig 4A,B**). Similar trends were observed for the stiffening traces obtained at the other speeds (**Fig. 4B**, **Fig S2**).

Two notable takeaways from these results are that active composites exhibit (1) emergent mechanical resistance at intermediate kinesin concentrations and (2) viscoelastic response to the application of local strains is heterogeneous, varying from stiffening to elastic to more viscous-dominated (i.e., yielding).

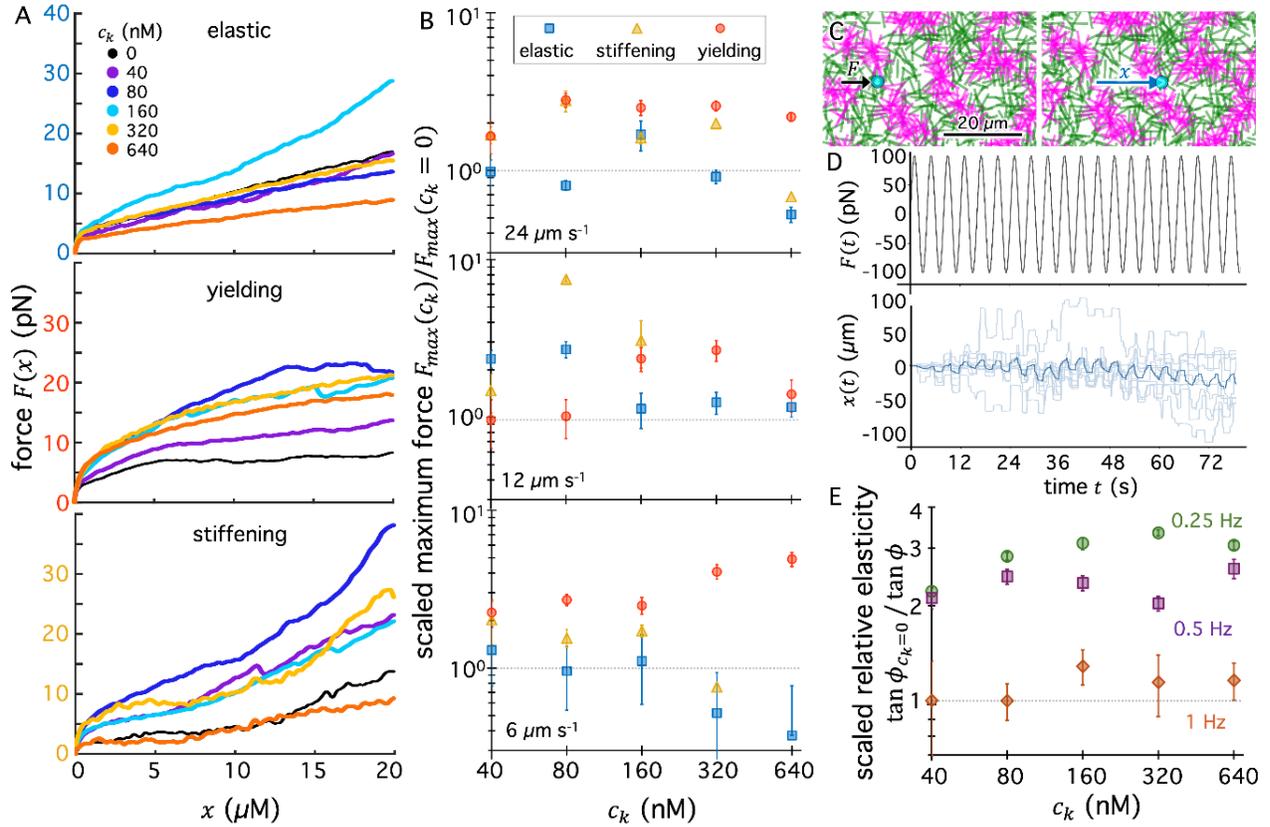

**Figure 4. Force response of kinesin-driven composites displays non-monotonic dependence on kinesin concentration.** (A) Average force $F(x)$ versus stage position $x$ measured in response to $v = 24$ μm s$^{-1}$ straining, for each kinesin concentration $c_k$, listed and color-coded according to the legend in the top panel. Force traces classified as elastic (top, blue y-axis), yielding (middle, red y-axis), and stiffening (bottom, gold y-axis) are averaged separately. Responses at $v = 6$ μm s$^{-1}$ and $v = 12$ μm s$^{-1}$ are shown in Fig. S2. (B) Maximum force reached during the strain $F_{max}$, determined from average force traces, and normalized by the corresponding value at $c_k = 0$ (denoted by the dashed horizontal line), for each class of response: elastic (blue squares), yielding (red circles), stiffening (gold triangles). Data plotted in the top panel correspond to the curves shown in A. Middle and bottom panels are for $v = 12$ μm s$^{-1}$ and $v = 6$ μm s$^{-1}$. Error bars correspond to standard error. (C) Sample simulated composite with embedded 1 μm particle (cyan) subject to force $F$ (black arrow, left) that displaces the particle a distance $x$ (blue arrow, right). (D) Simulated strains are sinusoidal with amplitude of $F_0 = 100$ pN and result in oscillatory particle displacements (light blue) which are averaged together (blue) to determine viscoelastic moduli $G'$ and $G''$ by evaluating the phase shift $\phi$ between $F$ and $x$. Sample data shown is for $c_k = 160$ nM and oscillation frequency of 0.25 Hz. (E) Scaled relative elasticity, computed as the inverse loss tangent $[\tan \phi]^{-1}$ normalized by the corresponding $c_k = 0$ value, indicated by the dashed horizontal line, versus kinesin concentration for strain frequencies of 0.25 Hz (green circles), 0.5 Hz (purple squares) and 1 Hz (red diamonds). Error bars correspond to standard deviation of bootstrapped ensembles.

To shed further light on these features, and assess the ability of our simulations to capture them, we introduced spherical probes into our simulated composites and imparted oscillatory forcing on them through the composite (**Fig 4C**). As described in Methods and SI, we measured the probe displacement resulting from oscillatory forcing with amplitude $F_0 = 100$ pN and frequencies $\omega = 0.25, 0.5,$ and 1 Hz, to determine the viscoelastic stress response (**Fig 4D**). We chose the force amplitude to achieve particle displacements comparable to our 20 μm experimental strain (Fig 4D), and frequencies to approximately match those in our experiments, considering $\omega = v/s$. Specifically, we measured the bead displacement amplitude and phase difference $\phi$ between the oscillation in applied force and resulting bead displacement to determine the elastic modulus $G'$ and viscous modulus $G''$ as a function of kinesin concentration (**Fig S3**). To quantify the relative elasticity of the composite we evaluated the inverse loss tangent $[\tan\phi]^{-1} = G'/G''$ which is increasingly >1 or <1 for more elastic-dominated or more viscous-dominated responses, respectively. We found that introducing kinesin into composites increased the relative elasticity for all frequencies (**Fig 4E**), and that peak elasticity was observed at intermediate kinesin concentrations for 0.25 Hz and 1 Hz. The 0.5 Hz data also showed a local maximum at intermediate kinesin concentrations but then increased again at $c_k = 640$ nM. These features align with our experimental results that display non-monotonic dependence of the force response on $c_k$ for most formulations and speeds; and highlight the importance of both elastic and viscous contributions to the force response. Collectively, these results demonstrate the emergent elasticity and force resistance that kinesin-driven de-mixing affords, which is optimized at intermediate kinesin concentrations.

To more quantitatively understand the tunable viscoelastic nature of the experimental force responses, and their underlying drivers, we use a mechano-equivalent circuit approach to model the ensembled-averaged responses.[46,47] Due to the relative infrequency of the stiffening responses, and the likelihood that that subset of traces is dominated by rare interactions of the particles with heterogeneous microstructures, we focused our analysis on the elastic and yielding responses only. We designed an equivalent circuit that consists of two Kelvin-Voigt elements in series (**Fig. 5A**, details in SI). The first element accounts for the composite network viscoelasticity, which is represented by a spring element with spring constant $\kappa$ to represent the network stiffness, and a dashpot element with drag coefficient $\gamma$ to represent viscous dissipation. A second Kelvin-Voigt element represents the effect of the optical trap stiffness $k_{OT}$ (which is known). We allowed the two elements to undergo relative deformation, and tracked the position of the center of the optical trap and the particle as $x_1$ and $x_2$, respectively. Here, $x_1 = vt$, where $v$ is the stage speed, $t$ is the elapsed time. To estimate the force response as a function of stage motion, we calculate $F = k_{trap}(x_1 - x_2)$. Assuming that at $t = 0$, $x_1 = x_2 \approx 0$, we established:

$$F(x_1) = \frac{k_{trap}\kappa}{k_{trap}+\kappa} \cdot x_1 - \gamma v k_{trap}\left(\frac{\kappa - k_{trap}}{(k_{trap}+\kappa)^2}\right)\left(1 - e^{-\left(\frac{k_{trap}+\kappa}{2\gamma}\right)x_1}\right) \tag{1}$$

We selected this model to capture the following phenomena: an initial elastic jump due to the re-equilibration of the particle within the optical trap as the stage begins to move, the transition to a second elastic regime as the particle engages with the composite network, and the presence of transient bonds that can dissipate stress and can be modeled via an effective viscosity (**Fig. 5B**). We use this approach to analyze each of the 6 kinesin concentrations at each of the 3 tested stage speeds.

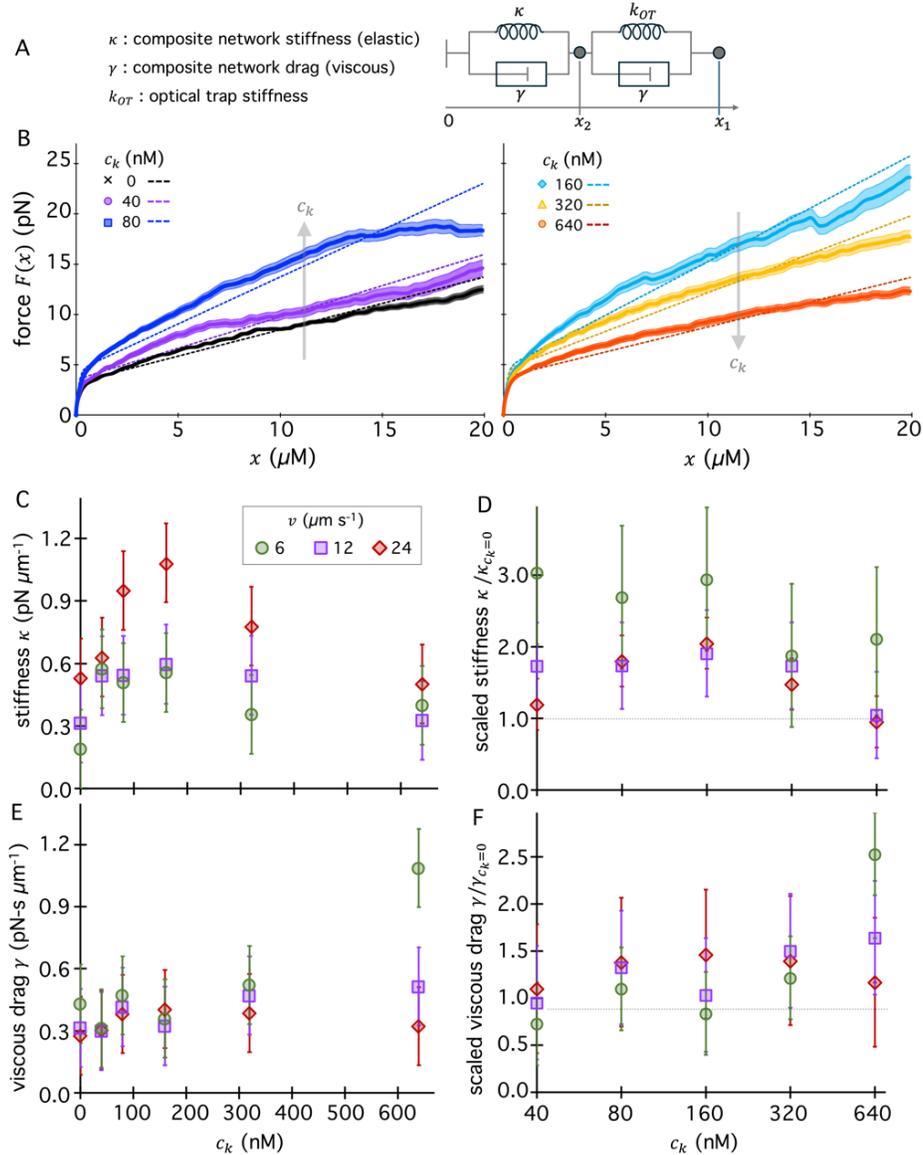

**Figure 5. Mechanical circuit model captures the viscoelastic behavior of elastic and yielding response classes and emergent stiffness at intermediate kinesin concentrations.** (A) Cartoon of mechanical circuit that models the force-displacement relationship for a bead pulled through a network by an optical trap with known trap stiffness $k_{OT}$. The composite network stiffness $\kappa$ and drag $\gamma$ are fit parameters in the model. (B) Force $F(x)$ versus stage position $x$, averaged across all elastic and yielding traces for each kinesin concentration $c_k$, listed and color-coded according to the legend, for $v = 24$ µm s$^{-1}$. Error bars denote standard error of the mean. Dashed lines are fits to the equation of motion for the mechanical circuit depicted in A. (C-F) Fit parameters $\kappa$ (C,D) and $\gamma$ (E,F) as a function of kinesin concentration $c_k$ for speeds $v = 6$ (green circles), 12 (purple squares) and 24 (red diamonds) µm s$^{-1}$, with error bars denoting 95% confidence intervals. Panels display (C,E) magnitudes for all kinesin concentrations on a linear scale and (D,F) values for $c_k > 0$ normalized by their corresponding $c_k = 0$ value and shown on a log scale.

For each speed, we found a nonmonotonic dependence of $\kappa$ on $c_k$, with the highest stiffness observed at $c_k = 160$ nM (**Fig. 5C,D**), consistent with our force analysis (**Fig. 4**); and structural assessments that showed a higher propensity for microtubule crosslinking and small-scale bundling at intermediate kinesin concentrations (**Fig 2**). Additionally, we found that the highest stiffnesses occurred at the fastest stage speeds, which may reflect the reduced ability for the network to relax on the timescale over which the strain is applied. Specifically, the presence of semiflexible actin filaments in the composites allows for actin bending modes to dissipate stress[44,48]. As described in SI Section S2, the predicted relaxation rate associated with actin bending in our composites is $\tau_b^{-1} \approx 25$ s$^{-1}$, which is comparable to the strain rate associated with our fastest speed, $\dot{\gamma} \simeq \frac{3v}{\sqrt{2}r_p} \simeq 23$ s$^{-1}$,[49] but faster than the two slower rates (~5.7 s$^{-1}$, ~11 s$^{-1}$). Thus, we may expect an increased likelihood of stress dissipation on the timescale of the slowest strain rate (i.e. at $v = 6$ μm s$^{-1}$) compared to the fastest (i.e. at $v = 24$ μm s$^{-1}$). Consistent with this understanding, we find that the viscous drag, which is a measure of stress dissipation within the composite network, is higher at slower speeds, particularly at the highest kinesin concentration (**Fig. 5E,F**).

**Hierarchical structural heterogeneity enables enhanced mechanical resistance for composites**

When we consider the mechanical results described above (**Figs 3-5**), in the context of the composite restructuring (**Fig. 1-2**), we see that the microtubule compaction and network de-mixing that causes dense small-scale clustering at intermediate kinesin concentrations also provides mechanical enhancement, as observed by the increase in both stiffness and maximum force. At higher kinesin concentrations the large-scale phase separation undermines and softens the elastic response. We now aim to quantitatively understand the relationship between de-mixing and structural heterogeneity and the non-monotonic dependence of local mechanics on kinesin concentration.

Our structural analysis shows varying degrees of clustering over a range of length scales from <10 μm to ~100 μm (**Fig 2**), which encompass the $s = 20$ μm displacement scale used in our optical tweezers experiments, as well as the forced bead displacements in simulations (**Fig 3D**). To determine the likelihood of observing structural heterogeneity within a local region that a moving particle perturbed and to better understand the extent of heterogeneity among different regions within an experimental field of view (FOV), we divided the epifluorescence videos we analyzed in **Fig 2** into 20 μm × 20 μm tiles or patches (**Fig 6A**). For each tile in the FOV, we defined a heterogeneity factor $\delta_I = \frac{\sigma_I}{\langle I \rangle}$ from the standard deviation $\sigma_I$ and mean $\langle I \rangle$ of pixel intensities $I$. To quantify heterogeneity at local ($< s$) and global ($> s$) scales for a given kinesin concentration, we computed the mean and standard deviation of $\delta_I$ across all tiles of all videos. The former and latter are measures of local heterogeneity, $h_I = \langle \delta_I \rangle$, and global heterogeneity, $H_I = \sigma(\delta_I)$, and their ratio $p_I = H_I/h_I$ is a measure of structural 'patchiness'. For reference, for a well-mixed system, there should be minimal global heterogeneity, i.e., all patches should be identical, so $p_I$ should tend to zero. For fractal-like systems, the heterogeneity would be scale invariant, yielding $p_I \approx 1$; and systems that are de-mixed on the scale of the patches, $p_I > 1$.

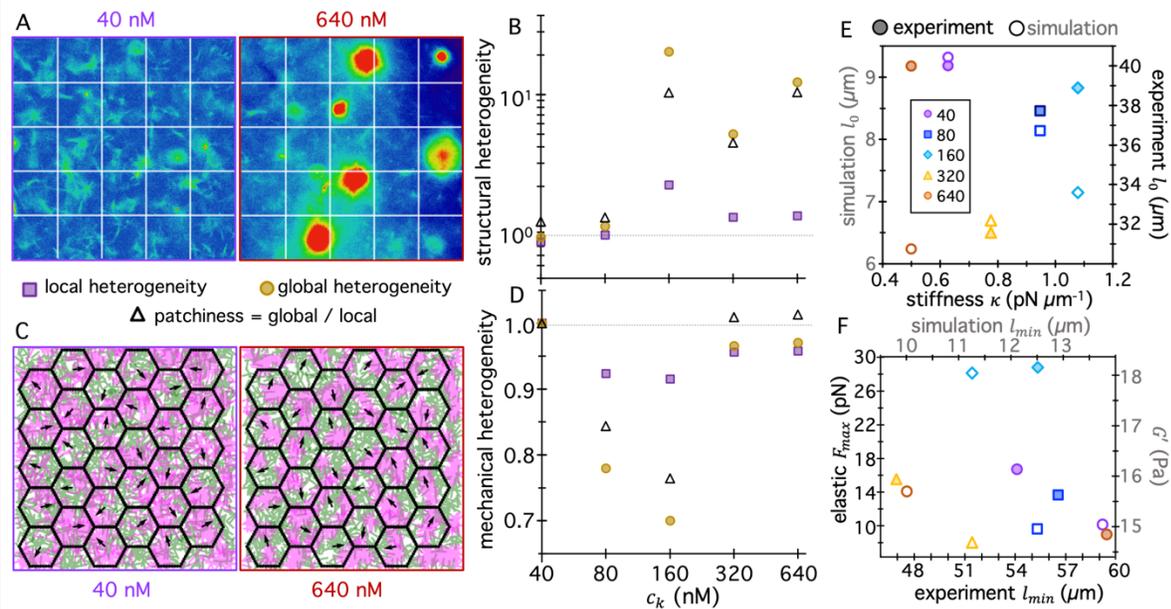

**Figure 6. Structural and mechanical heterogeneity and correlations underlying emergent stiffness of active composites**. (A) Sample epifluorescence images, with pixel intensity values shown in false-color from low (blue) to high (red), showing labelled actin and microtubules in composites with $c_k$ = 40 nM (left, purple border) and $c_k$ = 640 nM (right, red border). Each image was divided into a grid of 20 μm × 20 μm tiles (white lines) to compute local and global heterogeneity and patchiness parameters, $h_I, H_I$ and $p_I$. (B) Structural heterogeneity metrics $h_I$ (purple squares), $H_I$ (gold circles) and $p_I$ (open triangles), normalized by their corresponding $c_k$ = 0 nM value, denoted by the dashed horizontal line. (C) Sample snapshots of composites snapshots with $c_k$ = 40 nM (left, purple border) and $c_k$ = 640 nM (right, red border) with overlaid grid of 20 μm hexagonal tiles. The average force exerted on filaments within each tile, depicted as a black arrow, and the standard deviation of force values were used to compute mechanical heterogeneity metrics, $h_f, H_f$ and $p_f$ analogous to their structural counterparts. (D) Mechanical heterogeneity metrics $h_f$ (purple squares), $H_f$ (gold circles) and $p_f$ (open triangles), normalized by their corresponding $c_k$ = 0 nM value, denoted by the dashed horizontal line. (E,F) Correlation plots that display relationships between mechanical ($\kappa, F_{max}, G'$) and structural ($l_0, l_{min}$) parameters measured from experiments (filled symbols, black axis labels) and simulations (open symbols, grey axis labels). All experimental data shown is for 24 μm s$^{-1}$ strains, $F_{max}$ values are for the elastic traces, and simulated $G'$ data is for 1 Hz.

False-color images in **Fig 6A** depict the extent to which global heterogeneity and patchiness were enhanced and local heterogeneity was suppressed for $c_k$ = 640 nM compared to $c_k$ = 40 nM. We quantified this effect by plotting $h_I$, $H_I$ and $p_I$, normalized by their $c_k$ = 0 values, as functions of $c_k$ (**Fig 6B**). We observed a non-monotonic dependence on $c_k$ with peaks at $c_k$ = 160 nM, consistent with our prior observations. The maximum in $h_I$ is a likely indicator of local bundling of filaments that we expect to stiffen the network by stiffening its constituents, consistent with the increased maximum force and stiffness we measured (**Figs 3,4**). The maximum in $H_I$ is suggestive of larger scale de-mixing, and we found similarly high values for $c_k$ = 640 nM, as we expected based on our structural analysis (**Fig 2**) and visual inspection of the videos (**Fig 6A**). However, at this highest kinesin concentration, local heterogeneity dropped while the patchiness remained high. Together, these features suggest that larger scale aggregation and aster formation, with patches that are largely either filled by a cluster or devoid of microtubules, dominate the force response. This broken connectivity substantially weakens the network; and the prevalence of filament-poor patches compared to cluster-spanning patches tips the scales towards a softer mechanical response.

To verify this interpretation and couple structural and mechanical heterogeneity, we again turned to simulations, this time evaluating the distribution of forces exerted on filaments within 20 μm hexagonal tiles (**Fig 6C**), as fully described in the SI. We computed similar metrics to assess mechanical heterogeneity, replacing pixel intensity $I$ with force $f$ : $h_f = \langle \delta_f = \sigma_f / \langle f \rangle \rangle$ and $H_f = \sigma(\delta_f)$, and $p_I = H_I/h_I$. As shown in **Fig 6D**, in which we plot these metrics normalized by their values for the lowest kinesin concentration (i.e., $c_k = 40$ nM, metrics are not defined for $c_k = 0$ nM where $f = 0$), we observed strong non-monotonic dependence on $c_k$ consistent with our observations of the structural heterogeneity response shown in **Fig 6B**. However, and notably, all metrics were minimized for the $c_k = 160$ nM composite, with $H_I$ and $p$ being most strongly suppressed. This increased homogeneity of forces throughout the composite is consistent with the presence of a well-connected network of stiff (bundled) fibers that can both efficiently distribute stress and provide strong elastic resistance. At higher $c_k$ values, when de-mixing occurs at larger lengthscales, we observed a much higher patchiness of forces, signifying reduced mechanical connectivity, which is consistent with a weaker force response, loss of stiffness and enhanced yielding and viscous dissipation.

We have shown clear emergent elasticity of kinesin-driven composites in both experiments and simulations (**Fig 6E,F**). This response arises due to de-mixing of actin and microtubules and is a unique feature of the composite system (**Fig S4**). To summarize and provide further insight, we constructed correlation plots to depict the relationships between the structural correlation lengths obtained from SIA of experimental images with mechanical parameters of the network. Specifically, we compared the smaller structural lengthscale $l_0$ with stiffness $\kappa$ (**Fig 6E**); and the larger lengthscale $l_{min}$ with the experimental maximum force $F_{max}$ and simulated average elastic modulus $G'$ (**Fig 6F**). Across all metrics, we observed maximum elastic response for the $c_k = 160$ nM composite, as shown by the cyan diamonds being furthest to the right of **Fig 6E** and the top of **Fig 6F**. At intermediate stiffness values, we observed good agreement between experimental and simulated dependences of $l_0$ on $\kappa$, seen as the open and filled symbols closely aligning (**Fig 6E**). However, at the highest kinesin concentration, $c_k = 640$ nM, which has the lowest stiffness, simulations report the smallest $l_0$ among kinesin concentrations while experiments measured a maximal $l_0$. We believe that this distinction (**Fig 6E**), also seen in **Fig 2**, can be attributed to the reduced dimensionality and size of the 2D simulations which limits the ability for filaments to move and assemble into large clusters. Despite this simplification, we observe generally similar correlations between mechanical and structural properties for varying kinesin concentrations measured in experiments and simulations (**Fig 6F**). Comparing $F_{max}$ for the linear traces and the average simulated $G'$ values, and their dependences on their respective $l_{min}$ values, we find that simulated and experimental data points at a given kinesin concentration loosely cluster with one another, except for the $c_k = 640$ nM case for reasons described above. These general features confirm that our simulations are capturing the key physics of the material system, and highlight the importance of coupling between structure and mechanics to produce the emergent behavior.

**CONCLUSION**

We have designed and characterized in vitro composites of actin filaments and microtubules undergoing active restructuring by kinesin motor clusters that pull on microtubules, and found them to transition from interpenetrating networks to de-mixed microtubule-rich aggregates and actin-rich gas phases. Despite this de-mixing, composites maintain structural integrity, without fracturing or completely phase-separating, and achieve steady-states that maintain viscoelastic mechanical properties. We have discovered that this restructuring, seen in both experiments and simulations, leads to rich dependence of the mechanical response on kinesin concentration, including a surprising emergence of enhanced stiffness and elasticity at intermediate kinesin concentrations. This mechanical emergence is coupled to enhanced structural heterogeneity across lengthscales. Importantly, we previously observed non-monotonic dependence of mechanical stiffness on passive crosslinking of actin in cytoskeleton composites, suggesting this behavior may be a generalizable feature of crosslinking of one species of a composite. However, in these previous studies, there was no observable de-mixing, large-scale bundling or structural heterogeneity. Rather, the non-monotonic dependence was a result of modest microscale variations in mesh sizes and fiber stiffnesses. Here, our experiments and simulations demonstrate the importance of hierarchical structural and mechanical heterogeneity in sculpting the mechanical behavior, which we rationalize as a direct result of the internal stress generated by kinesin motors. The distribution of motor-generated stresses measured in simulations mirrors that of the structural heterogeneity in experiments. Moreover, the stiffening behavior, a unique feature not previously reported in similar passive or active composites[28,35], also may indicate motor-generated pre-stress and densification that suppress filament bending and non-affine deformations that dissipate stress, thereby promoting a stiffening response.

This interplay between structure and mechanics is likely critical to the multifunctionality of the cytoskeleton that allows for wide-ranging mechanical processes and dynamically sculpts mechanical properties in response to environmental cues and the needs of the cell. Our results and models shed important light on how to engineer and tune composite systems to exhibit emergent mechanics through independent tuning of elastic and viscous contributions of the composite constituents.

**MATERIALS AND METHODS**

**Protein Preparation:** Rabbit skeletal actin (Cytoskeleton, Inc. AKL99) was reconstituted to 2 mg mL$^{-1}$ in 5 mM Tris-HCl (pH 8.0), 0.2 mM CaCl$_2$, 0.2 mM ATP, 5% (w/v) sucrose, and 1% (w/v) dextran. Porcine brain tubulin (Cytoskeleton, Inc. T240) and HiLyte488-labeled porcine brain tubulin (Cytoskeleton, Inc. TL488M-A) were reconstituted to 5 mg mL$^{-1}$ with 80 mM PIPES (pH 6.9), 2 mM MgCl$_2$, 0.5 mM EGTA, and 1 mM GTP. All cytoskeleton proteins were flash frozen single-use aliquots and stored at -80ºC.

Biotinylated kinesin-401[21,50] was expressed in Rosetta (DE3) pLysS competent E. coli cells (ThermoFisher), purified, and flash-frozen into single-use aliquots, as described previously[33]. To prepare force-generating kinesin clusters, kinesin-401 dimers were incubated with NeutrAvidin (ThermoFisher) at a 2:1 ratio in PEM-100 buffer (100 mM PIPES, 2 mM $MgCl_2$, 2 mM EGTA) supplemented with 4 µM DTT for 30 min at 4°C. Clusters were prepared fresh and used within 24 hrs.

**Composite Sample Preparation:** Composites of actin filaments and microtubules at a 45:55 molar ratio, were polymerized by combining 1.35 µM actin monomers, 1.55 µM tubulin dimers, and 0.1 µM HiLyte488 tubulin dimers in PEM-100 (100 mM PIPES, 2 mM $MgCl_2$, 2 mM EGTA) supplemented with 0.1% Tween, 10 mM ATP, 4 mM GTP, 5 µM Taxol, 1.08 µM phalloidin, and 0.27 µM ActiStain488 phalloidin (Cytoskeleton, Inc. PHDG1-A) and incubating for 1 hr at 37°C. The fluorophores for both microtubules and actin were chosen to be spectrally similar to allow both filaments to be visible in a single field-of-view with the same excitation/emission filter combination, a requirement due to the fact that the epifluorescence microscope outfitted with our optical tweezers can only accommodate a single fluorescence channel at a time. For optical tweezers experiments, 0.02% (v/v) of 4.5 µm diameter carboxylated microspheres (Polysciences, Inc.), coated with BSA to inhibit non-specific interactions with the network, were added[28]. For two-color confocal microscopy measurements, HiLyte488 tubulin dimers were replaced with rhodamine tubulin dimers (Cytoskeleton, Inc. TL590M) to allow for separate imaging of actin and microtubules in different channels of the confocal microscope (see below for additional details).

Following polymerization, and immediately prior to experiments, an oxygen scavenging system (45 µg $mL^{-1}$ glucose, 43 µg $mL^{-1}$ glucose oxidase, 7 µg $mL^{-1}$ catalase, 0.005% β-mercaptoethanol) was added to reduce photobleaching, followed by kinesin clusters at final kinesin concentrations of $c_k$ = 0, 40, 80, 160, 320, and 640 nM.

For both optical tweezers and confocal experiments, the final sample was gently flowed into a sample chamber made from a glass slide and coverslip separated by ~100 µm of double-stick tape to accommodate ~10 µL. Both the glass slide and coverslip of the chamber were passivated with BSA prior to flowing in the sample. The chamber was sealed with UV curable glue to prevent sample leakage and evaporation. This process completed ~5 mins after the addition of kinesin to the sample, and the sealed sample was incubated for an additional 25 mins to allow for motor-driven restructuring prior to measurements.

The composite mesh size $\xi$ is determined from the mesh sizes for the actin and microtubule networks comprising the composite, $\xi_A \simeq 1.46 c_A^{-1/2} \simeq 1.26$ µm and $\xi_M \simeq 2.68\ c_T^{-1/2} \simeq 2.15$ µm, where $c_A$ and $c_T$ are the molarities of actin and tubulin, via the relation $\xi \simeq (\xi_A^3 + \xi_T^3)^{-1/3}$,[44] yielding a composite mesh size of $\xi \simeq 1.19$ µm. The ratio of kinesin clusters to tubulin $R = \frac{c_k}{4 c_T} =$ 0, 0.006, 0.012, 0.024, 0.048 and 0.097 for $c_k$ = 0 – 640 nM, where the 4 accounts for the ~4 kinesins (two dimers) per motor cluster.

**Optical Tweezers Microrheology (OTM)**: OTM experiments were performed using an optical trap formed by a 1064 nm Nd:YAG fiber laser (Manlight), focused with a 60× 1.4 NA objective (Olympus), and custom-built around an Olympus IX71 epifluorescence microscope[41,51]. The force was measured using a position-sensing detector (Pacific Silicon Sensor) to record the deflection of the trapping laser, which over the operating range used within the reported experiments, is proportional to the force acting on the trapped microsphere. The proportionality constant that provides the trap stiffness $k_{OT}$ was determined to be $k_{OT} \simeq 68$ pN/µm using the Stokes drag method[40,41,51]. Imaging of the labeled filaments and probes was achieved using a broadband LED source (XCITE) with 488nm/525nm excitation/emission filters and a Hamamatsu ORCA-Flash 4.0LT CMOS camera with a 1024 x 1024 square-pixel field-of-view and frame rate of 20 s$^{-1}$.

For each force measurement, an optically trapped microsphere was dragged back and forth in the $\pm x$-direction through the sample in a sawtooth pattern using a nanopositioning piezoelectric stage (Mad City Labs) that moves the sample chamber relative to the fixed trap (**Fig 3a**). The distance the probe was moved in each half-cycle (i.e., the strain amplitude) and total perturbation time was fixed at $s = 20$ μm and $t_f = 50$ s for all measurements. The stage position and laser deflection were recorded at 20 kHz, and the stage position was updated at 400 Hz using custom-written National Instruments LabVIEW programs. Measurements were performed at 3 different speeds for each kinesin concentration: $v$ = 6, 12, 24 μm s$^{-1}$. Prior to each strain and force measurement, an image of the labeled filaments surrounding the probe in a 202 μm x 135 μm (900 x 600 pixel) area centered at the center of the strain path was captured. The same protocol was repeated but using a piezoelectric mirror to move the trapped bead relative to the sample (keeping the sample chamber fixed), and recording 1000-frame videos of the labeled filaments. These videos were used in the image analysis presented in **Figs 2** and **6**.

For each ($v, c_k$) combination, measurements were repeated using multiple beads located in different regions of the sample chamber, which were separated by ≥100 μm. This suite of measurements for each condition was then repeated for three different samples for a total of ≥24 beads per condition. While we performed cyclic straining for each measurement, in which the same probe was repeatedly pulled through the material, we found that a significant number of probes were lost on the return after the initial pull, and that for those that were not lost, the measured force ranges for subsequent pulls were considerably smaller, suggesting that the initial pull caused network damage or plastic deformation. Thus, data presented in **Figs 3-6** are solely for the initial loading cycle.

Post-acquisition analysis of measured force $F(x,t)$ (Figs 3-6) was performed using custom-written MATLAB scripts. Each 50 s measurement was divided into individual cycles and the second half of each cycle, when the probe is moving in the $-x$ direction, was removed. The last 1% of the forward cycle (0.2 μm) was also removed to avoid artifacts that arise from attempting to instantaneously switch the stage motion from $+v$ to $-v$ (the stage response rate is 400 Hz). Average data shown for each condition represent averages over all valid trials and error bars represent standard error. Trials were considered invalid if the bead was pulled out of the trap.

To characterize the composite structure, we performed Spatial Image Autocorrelation (SIA) analysis[39,51] on each frame of each of the 1000-frame videos described above. SIA measures the correlation in intensity $g_I$ of two pixels in an image as a function of separation distance $r$. Autocorrelation curves $g_I(r)$ were generated by taking the fast Fourier transform of the image $F(I)$, multiplying by its complex conjugate, applying an inverse Fourier transform $F^{-1}$, and normalizing by the squared intensity: $g_I(r) = \frac{F^{-1}(|F(I(r))|^2)}{[I(r)]^2}$. To determine the effect of motor activity on the structure we subtract $g_I(r)$ for the no-motor case ($c_k$ =0) from $g_I(r)$ for each kinesin concentration: $\Delta g_I(r, c_k) = g_I(r, c_k) - g_I(r, 0)$, which we show in **Fig 2**. We observed no dependence on strain speed so the data shown is the average and standard error across all frames of all videos for a given kinesin concentration $c_k$.

**Fluorescence Confocal Microscopy:** To determine the unperturbed composite structure and dynamics, videos of composites with distinctly-labeled actin and microtubules were recorded using a Nikon A1R laser scanning confocal microscope with a 60× 1.4 NA oil-immersion objective (Nikon), 488 nm laser with 488/525 nm excitation/emission filters (to excite/image actin), and 561 nm laser with 565/591 nm excitation/emission filters (to excite/image microtubules) (**Fig 1**). For each kinesin concentration, four time-series (videos) of 512 × 512 square-pixel (213 µm × 213 µm) images were collected at 1.86 fps for a total of 1116 frames (10 mins). Videos were collected at 10, 20, 30 and 40 mins after the addition of kinesin motors, with each video collected in a different field of view separated by >500 µm. We observed no dependence of the composite structure on the time that the video was acquired, indicating the motor activity is largely halted after 10 mins. All videos include two channels that separate the actin and microtubule signals such that they can be processed separately and compared.

**Computational Model**: To predict the restructuring of the composites due to motor activity, we develop a minimal model, fully described in Supplementary Information, that captures the key components of the composites. We define the available space as a hexagonal grid with periodic boundary conditions. Each grid point can be occupied by a microtubule or actin filament center or can be empty. The filaments can interact with neighbors within reach, via 1) motor-generated forces that can either pull the interacting filaments towards each other or push them away from each other; and 2) motor crosslinks that increase the friction forces on the interacting filaments and allow forces to be transmitted through crosslinked filament clusters. The movement of a filament center to a neighboring grid point within a small temporal time step is then a stochastic event whose probability can be calculated using the transition rate based on the first passage times.

We purposefully choose a minimal approach to capture the dynamics to shed light on the competing factors of activity and friction. Our model assumes a single length for all filaments of $l = 5$ µm while in experiments actin and microtubules assume distributions of lengths ($l_A \simeq 4 \pm 3$ µm and $l_M \simeq 8 \pm 4$ µm)[44]. We treat all filaments as rigid rods while actin in experiments is semiflexible with a persistence length of ~17 µm. Our simulations are in 2D while experimental composites span 3D.

We implement our model on a 100 µm x 100 µm 2D space with a hexagonal lattice, where the lattice spacing is 1.25 µm. Each 5-µm long filament interacts with other filaments located within 4 grid points in all directions. Initially, each lattice point is either occupied with a microtubule center, an actin filament center, or is left empty using probabilities matching the average volume fraction occupied by these elements. The movement of the filaments is simulated in each iteration by calculating the likelihood of each possible movement, $p_{ij}$ for all grid points i and j, where at least one of them contains a filament center, and randomly picking one of these movements to occur based on these probabilities. The simulation is run for $T_S = 5$ minutes, which we find is sufficient to reach quasi-steady state. The model calculations and simulations are coded in python and the scripts are available on GitHUB[52]. A cartoon depiction of the model is shown in Figs 1 and S1 and numerical values for all model parameters are included in Table S1.

To quantify the degree of clustering and segregation of the different filaments, we compute the filament pair distribution function $g_{AM}(r,T)$ where the subscripts $A$ and $M$ represent the filament type, either actin or microtubules, and which gives the probability of finding a filament (actin or microtubule) a radial distance $r$ from any other

like filament:

$$g_{AA}(r) = \langle \frac{N_A(r)}{f_A N(r)} \rangle, \; g_{MM}(r) = \langle \frac{N_M(r)}{f_M N(r)} \rangle \tag{2}$$

or unlike filament

$$g_{MA}(r) = \langle \frac{N_A(r)}{f_A N(r)} \rangle, g_{AM}(r) = \langle \frac{N_M(r)}{f_M N(r)} \rangle \tag{3}$$

where $N_A(r)$ is the number of neighboring filaments of type $A$ a distance $r$ from a specific filament, $f_A$ is the volume fraction of filament $A$ in the simulation space, and $N(r)$ is the maximum number of possible neighbors a distance $r$ from the specific filament. An increase in $g_{AA}(r)$ above 1 indicates clustering of like filaments, and a decrease in $g_{MA}(r)$ below 1 indicates segregation of unlike filaments. As with experimental SIA data, we subtract the distribution for the no-motor case from that for each kinesin concentration to yield: $\Delta g_{AA}(r, c_k) = g_{AA}(r, c_k) - g_{AA}(r, 0)$, which we plot in Fig 2a-c. We perform correlation analysis up to $r = 25$ μm which we found sufficient to capture most of the correlation decay.

We calculate the storage modulus $G'$ and loss modulus $G''$ of each quasi-steady state composite by embedding a spherical bead of radius $r_{sph} = 0.625$ μm into the in silico composite and applying a sinusoidal force on the bead with amplitude $F_0 = 100$ pN and oscillation frequencies $\omega = 0.25, 0.5, 1$ Hz for 20 full periods. We measure the resulting displacement of the bead through the composite, and use fast Fourier transform analysis to compute the magnitude $x$ and phase angle $\varphi_x$ of the displacement at the chosen forcing frequency $\omega$. By combining this data with the known amplitude $F_0$ and phase angle $\varphi_F$ of the applied force, we calculate the viscoelastic moduli as

$$G' = \frac{F_0}{x} * \cos\phi * \frac{1}{r_{sph}} \tag{4}$$

$$G'' = \frac{F_0}{x} * \sin\phi * \frac{1}{r_{sph}} \tag{5}$$

where $\phi = \varphi_F - \varphi_x$ is the phase difference. We quantify the relative elasticity of the response by evaluating the inverse loss tangent $[\tan\phi]^{-1} = G'/G''$ which is greater or less than 1 for elastic-dominated and viscous-dominated responses, respectively. Each data point shown in **Figs 3, S2, S3** is the average over 10 beads in each of 3 replicate samples and normalized by the corresponding value for $c_k = 0$. Error bars are computed by bootstrapping of 10 random subsets of the data as described in SI.

## ACKNOWLEDGEMENTS

We acknowledge funding from the US National Science Foundation DMREF program through following grants: NSF DMR-2119663 (to RMRA), NSF DMR-2118497 (to MTV), NSF DMR-2118403 (to JLR); the US Department of Defense, Army Research Office W911NF-23-1-0329 (to PK); the US National Institutes of Health 2R15GM123420-02 (to RMRA, JYS); and the US National Science Foundation DMR-2203791 (to JYS). We acknowledge the contributions of Daisy Achiriloaie in collecting experimental data, Moumita Das and Ryan McGorty for insightful discussions, Karthik Peddireddy for acquisition software design, and Dylan McCluskey for assistance in preprocessing the force data.

## AUTHOR CONTRIBUTIONS

JS and RMRA conceived and designed the research. JS, CG, KM, MS, PK, MTV, and RMRA performed research and analyzed data. RMRA, CG, PK, and MTV prepared figures and wrote the manuscript. PK and RMRA supervised the research. JLR prepared reagents. All authors interpreted data and edited the manuscript.

## DATA AVAILABILITY

All data presented in the main text or SI is available upon request.

**Kinesin-driven de-mixing of cytoskeleton composites drives emergent mechanical properties**

Janet Sheung[1], Christopher Gunter[2], Katarina Matic[3], Mehrzad Sasanpour[3], Jennifer L. Ross[4], Parag Katira[2], Megan T. Valentine[5], Rae M. Robertson-Anderson[3*]

[1]Department of Natural Sciences, Scripps and Pitzer Colleges, Claremont, CA, United States; W. M. Keck Science Department, Claremont Mc College, Claremont, CA 91711, USA
[2]Department of Mechanical Engineering, San Diego State University, San Diego, CA 92182, USA
[3]Department of Physics and Biophysics, University of San Diego, San Diego, CA 92110, USA
[4]Department of Physics, Syracuse University, Syracuse, NY 13244, USA
[5]Department of Mechanical Engineering, University of California, Santa Barbara, Santa Barbara CA 93106, USA

*Corresponding Author: randerson@sandiego.edu

**Supplemental Information**

**Section S1. Computational Model**

**Table S1: Parameters used in mathematical model and simulations**

**Figure S1: Sample plot showing simulation mechanics**

**Section S2. Predicted Relaxation Timescales**

**Section S3. Mechanical Circuit Model**

**Figure S2. Average force traces classified by response type for strains of speed $v = $ 6 µm s$^{-1}$ (top) and $v = $ 12 µm s$^{-1}$ (bottom).**

**Figure S3. Viscoelastic moduli determined from simulated oscillatory forcing of a bead through active cytoskeletal composites at varying oscillation frequencies.**

**Figure S4. Viscoelastic moduli determined from simulated oscillatory forcing of a bead through an active microtubule network without actin present.**

**Figure S5. Mechanical circuit model captures the viscoelastic behaviour of elastic and yielding response classes.**

**Figure S6. Ensemble of force traces comprising averages shown in Figure 4**

**Supplementary References**

**Section S1.**

**Computational Model:** To predict the restructuring of the composites due to motor activity, we developed a minimal model that represents the principal components of the composites, and captures their key dynamics. We defined the available space as a hexagonal grid with periodic boundary conditions. Within the model, each grid point can be occupied by an actin or microtubule filament center or can be empty. The filaments can interact with neighboring filaments within reach, via 1) motor-generated forces that can either pull the interacting filaments towards each other or push them away from each other; and 2) crosslinks that increase the friction forces on the interacting filaments and allow forces to be transmitted through crosslinked filament clusters. We implemented a parameter that designates a portion of the motors as active, in which case they exert forces on interacting filaments, and a portion as passive, in which case they crosslink interacting filaments together without exerting force. The movement of a filament center to a neighboring grid point within a small temporal time step is then a stochastic event whose probability can be calculated using the transition rate based on the first passage time. This probability is given by

$$p_{ij} = \frac{k_{ij}}{\Sigma k_{ij}} \qquad \text{S1,}$$

where $k_{ij}$ is the rate constant describing the transition rate of a filament at grid location $i$ in the direction of grid location $j$. This rate constant is calculated as

$$k_{ij} = \frac{k_0}{f_i} + \frac{v_i \cdot \delta}{s} \qquad \text{S2,}$$

Where $k_0$ is the free filament transition rate, $f_i$ is the friction factor associated with the filament, $v_i$ is the velocity of the filament, $\delta$ is a unit vector in the direction of motion from grid location $i$ to grid location $j$, and $s$ is the distance between the two grid locations.

The friction factor of a single filament is given by

$$f_i = 1 + \frac{\gamma_{mot} \ast (N_{mot,act} + N_{mot,pas}) \ast N_{fil}}{\gamma_{fil}} \qquad \text{S3,}$$

where $\gamma_{mot}$ is the friction coefficient of a single motor protein ($\gamma_k$ for a kinesin motor), $N_{mot,act}$ and $N_{mot,pas}$ are the number of active and passive motor proteins per filament (randomly selected from a Poisson distribution based on the mean value), $N_{fil}$ is the number of filaments within an interaction distance, and $\gamma_{fil}$ is the friction coefficient of a single filament ($\gamma_M$ for a microtubule).

The velocity term is given by

$$v_i \cdot \delta = \frac{F_i \cdot \delta}{\gamma_{fil} + \gamma_{mot} \ast (N_{mot,act} + N_{mot,pas}) \ast N_{fil}} \qquad \text{S4,}$$

where $F_i$ is the net force generated by the motors between filament $i$ and all same-type filaments within an interaction distance. This net force is given by

$$F_i = \Sigma F_{ij} \qquad \text{S5,}$$

where $F_{ij}$ is given by the force per motor ($F_k$ for kinesin) times the number of active motors per filament ($N_{k,act}$), which is randomly selected from a Poisson distribution based on the mean value. The direction of $F_{ij}$ is along the line joining the two filament centers and can be attractive or repulsive. This results in a filament orientation which is aligned with the direction of the vector sum of the forces exerted by all filaments of the same type within an interaction distance.

The movement of a filament center to a neighboring grid point occupied by another filament center is restricted sterically and can be only accomplished if the two filaments exchange positions. Thus, in such a scenario, the movement probability of filament $i$ to a neighboring grid point containing filament $j$'s center is given by

$$p_{i \leftrightarrow j} = p_{j \leftrightarrow i} = p_{i \rightarrow j} \times p_{j \rightarrow i} \qquad \text{S6,}$$

which is the same for the filament at grid point $j$ exchanging its location with filament at $i$.

In the same spirit, the movement of a filament from grid point $i$ to a neighboring empty grid point $j$ is given by

$$p_{i \leftrightarrow j} = p_{i \rightarrow j} \times 1 \qquad \text{S7.}$$

We purposefully chose a minimal approach to capture the composite dynamics to shed light on the competing factors of motor activity and friction from crosslinkers. Within this simplified approach, out model assumed a single length for all filaments, while in experiments actin and microtubules display a distribution of lengths. We treated all filaments as rigid rods while actin in experiments is semiflexible with a persistence length of ~17 µm. Our simulations were executed in 2D while experimental composites span 3D space.

We implemented our model on a 100 µm x 100 µm 2D space with a hexagonal lattice, with a lattice spacing of 1.25 µm. Each microtubule or actin filament was assumed to be 5 µm in length, such that each filament interacted with other filaments located within 4 grid points in all directions. Initially, each lattice point was either occupied with a microtubule center, an actin filament center, or was left empty using probabilities matching the average volume fraction occupied by these elements. The movement of the filaments was simulated in each iteration by calculating the likelihood of each possible movement, $p_{ij}$ for all grid points $i$ and $j$, where at least one of them contains a filament center, and randomly picking one of these movements to occur based on these probabilities. Since each movement occurred over a timescale of $\frac{1}{k_{ij}}$, the effective time progression for a single movement to have occurred was approximated by selecting a random value from an exponential distribution with a mean of $\frac{1}{\sum k_{ij}}$ ($i \neq j$, and at least $i$ or $j$ was occupied by a filament center) at each iteration step, following the Gillespie algorithm. Thus, $\Delta t$, the time progression, at each iteration step was dynamically adjusted to match the ongoing system dynamics. The simulation was run for $T_S = 5$ minutes, which we find to be sufficient to reach quasi-steady state. Specifically, running the simulation for $0.5T_S$, $0.8T_S$, $T_S$, and $1.2T_S$ iterations, we observed insignificant change in the filament distributions for $\geq 0.8T_S$. Additionally, we observed that the sum of all rate constants for each filament type reaches a quasi-stable non-zero value, implying quasi-steady state kinetics. The model calculations and simulations were coded in Python and the scripts are available on GitHUB (https://github.com/compactmatterlab/active-filament-networks/2024)[1]. A schematic depiction of the model is shown in Fig S1 and numerical values for all model parameters are included in Table S1.

*Structural analysis:* To quantify the degree of clustering and segregation of the different filaments, we computed the filament pair distribution function $g_{ij}(r,T)$ where the subscripts $i$ and $j$ represent the filament type, either actin (A) or microtubules (M), and which gives the probability of finding a filament (actin or microtubule) a radial distance $r$ from any other filament. The probability of finding a filament (actin or microtubule) a radial distance $r$ from any other like filament is given by:

$$g_{AA}(r) = \langle \frac{N_A(r)}{f_A N(r)} \rangle, g_{MM}(r) = \langle \frac{N_M(r)}{f_M N(r)} \rangle \qquad \text{S8,}$$

or unlike filament

$$g_{MA}(r) = \langle \frac{N_A(r)}{f_A N(r)} \rangle, g_{AM}(r) = \langle \frac{N_M(r)}{f_M N(r)} \rangle \qquad \text{S9,}$$

where $N_A(r)$ is the number of neighboring filaments of type $A$ a distance $r$ from a specific filament, $f_A$ is the volume fraction of filament $A$ in the simulation space, and $N(r)$ is the maximum number of possible neighbors a distance $r$ from the specific filament. An increase in $g_{AA}(r)$ above 1 indicates clustering of like filaments, and a decrease in $g_{MA}(r)$ below 1 indicates segregation of unlike filaments. As with experimental SIA data, we subtracted the distribution for the no-motor case from that obtained for each kinesin concentration to yield: $\Delta g_{AA}(r, c_k) = g_{AA}(r, c_k) - g_{AA}(r, 0)$, which we plotted in Fig 2a-c. We performed correlation analysis up to $r =$ 25 μm which we found sufficient to capture most of the correlation decay. Each scenario was simulated for 10 iterations and the results were combined to determine the average value and standard error.

*Viscoelastic response:* We calculated the storage modulus $G'$ and loss modulus $G''$ of each quasi-steady state composite by embedding a spherical bead of radius $r_{sph} = 0.625$ μm into the *in-silico* composite and applying a sinusoidal force on the bead with amplitude $F_0 = 100$ pN and oscillation frequencies $\omega = 0.25, 0.5, 1$ Hz for 20 full periods. We measured the resulting displacement of the bead through the composite, and used fast Fourier transform analysis to compute the magnitude $x$ and phase angle $\varphi_x$ of the displacement at the chosen forcing frequency $\omega$. By combining this data with the known amplitude $F_0$ and phase angle $\varphi_F$ of the applied force, we calculated the viscoelastic moduli as

$$G' = \frac{F_0}{x} * \cos\phi * \frac{1}{r_{sph}} \qquad \text{S10,}$$

$$G'' = \frac{F_0}{x} * \sin\phi * \frac{1}{r_{sph}} \qquad \text{S11,}$$

where $\phi = \varphi_F - \varphi_x$ is the phase difference. We quantified the relative elasticity of the response by evaluating the inverse loss tangent $[\tan\phi]^{-1} = G'/G''$ which is greater or less than 1 for elastic-dominated and viscous-dominated responses, respectively. Each data point shown in Figs 3, S3,S4 represented an average over 10 beads in each of 3 replicate samples.

We performed this analysis at 3 different frequencies (0.25 Hz, 0.5 Hz, and 1 Hz) and applied the sinusoidal force to the spherical bead for 20 full periods. Each scenario was simulated for 30 iterations and a bootstrapping technique was used to determine the average value and standard error. In this technique, a sample of 10 iterations is randomly selected from the pool of 30 iterations, and the results are combined to determine an average value for that sample. This process was performed 10 times, and the results from the 10 samples were then used to determine the final average value and standard error.

*Force heterogeneity:* We calculated the local and global heterogeneity of the forces exerted on filaments throughout each composite by first superimposing a grid of 30 hexagonal tiles, each with a long diagonal of 20 μm, over the composite network. For each tile, we evaluated the net force $f$ acting at each grid location within the tile, and computed the mean $\langle f \rangle$ and standard deviation $\sigma_f$ of the force ensemble, from which we determined the mechanical heterogeneity within each tile $\delta_f = \frac{\sigma_f}{\langle f \rangle}$. We computed the local and global mechanical heterogeneity factors, $h_f$ and $H_f$, as the mean and standard deviation of the 30 individual $\delta_f$ values: $h_f = \langle \delta_I \rangle$ and $H_I = \sigma(\delta_I)$.

|  | Description | Value | Reference |
|---|---|---|---|
| Total Grid Size |  | 100 μm x 100 μm |  |
| % actin filaments | % of 2D space taken up by actin filament | 25% | experimental |
| % microtubules | % of 2D space taken up by microtubules | 30% | experimental |
| Grid spacing ($l$) | Distance between grid locations | 1.25 μm |  |
| Filament length | Length of each actin filament and microtubule | 5 μm | experimental |
| $F_k$ | Force generated per kinesin motor | 6 pN | [2] |
| $N_{k,act}$ | Number of active kinesin motors per microtubule-microtubule interaction | variable | experimental |
| $N_{k,pas}$ | Number of passive kinesin motors per microtubule-microtubule interaction | variable | experimental |
| $\gamma_A$ | Viscous drag on an actin filament | 0.005 pN*ms/nm | [3] |
| $\gamma_M$ | Viscous drag on a microtubule filament | 0.01 pN*ms/nm | [4] |
| $\gamma_k$ | Viscous drag on the filament due to single kinesin motor binding | 6 pN*ms/nm | [4] |
| $k_{0,A}$ | Free filament transition rate for actin filament | 1.28 s$^{-1}$ | calculated using first passage time to a neighboring grid point |
| $k_{0,M}$ | Free filament transition rate for microtubule | 0.64 s$^{-1}$ | Calculated using first passage time to a neighboring grid point |
| $F_0$ | Force applied to spherical bead | 100 pN | experimental |
| $r_{sph}$ | Radius of spherical bead | 0.625 μm | lattice resolution |

**Table S1: Parameters used in mathematical model and simulations.** Specific numerical values of parameters were chosen to match experimental conditions, including the concentrations of actin, microtubules, and kinesin. Values for motor forces and viscous drag terms were based on literature values, as specified.

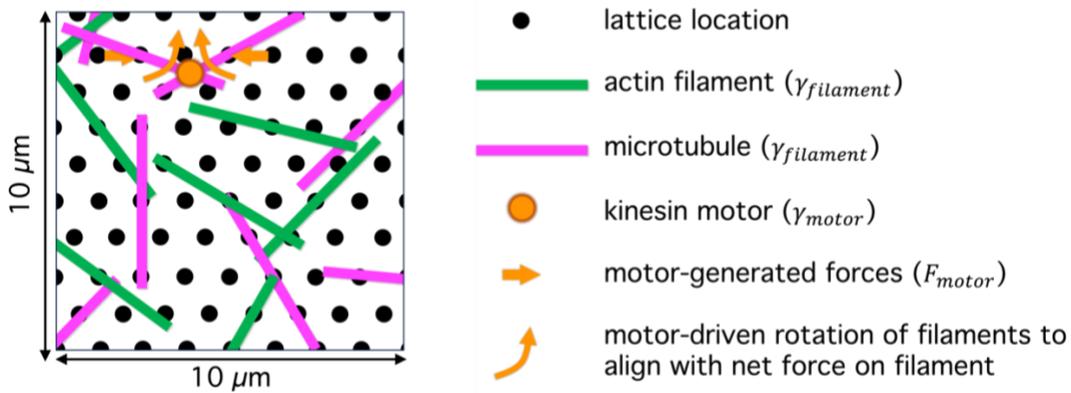

**Figure S1: Sample plot showing simulation mechanics.** The actin filaments and microtubules exist on a lattice of grid points. There is a drag ($\gamma_{filament}$) associated with their movement. Motor proteins exert forces which drive movement of the filaments ($F_{motor}$) but also exert drag ($\gamma_{motor}$).

**Section 2. Predicted Relaxation Timescales**

Several predicted relaxation timescales corresponding to various mechanisms have been shown to play a role in the force response of cytoskeleton networks[5]. The shortest relaxation timescale is that over which hydrodynamic interactions (HI) between filaments become important, termed the mesh time $\tau_\xi \approx \epsilon \xi^4 l_p^{-1}$ where $\xi$ is the mesh size, $l_p$ the persistence length and $\epsilon = \zeta/k_B T$ is the friction term, which we assume to be of order ~1 s µm$^{-3}$.[6,7] This expression yields $\tau_{\xi,A} \approx 118$ ms and $\tau_{\xi,M} \approx 2$ ms for the actin filaments and microtubules in the composites which equate to relaxation rates of $\tau_{\xi,A}^{-1} \approx 8.5$ s$^{-1}$ and $\tau_{\xi,M}^{-1} \approx 500$ s$^{-1}$. We can compare these values to our experimental strain rates, $\dot{\gamma} \simeq 3v/\sqrt{2}r_p \simeq 5.7, 11.3, 22.6$ s$^{-1}$, where $v$ and $r_p$ are the speed and radius of the optically trapped probe[8]. The fact that the mesh rates for both filaments are faster than or similar to all strain rates suggest that HI are important across the entire range of strains. In other words, because $\dot{\gamma} \lesssim \tau_\xi^{-1}$ the filaments have enough time to interact through HI over the timescale over which the strain is applied. In cases in which $\dot{\gamma} > \tau_\xi^{-1}$, the response is expected to be that of single non-interacting filaments. Another important timescale for semiflexible actin is the timescale associated with bending $\tau_b \approx [4\pi\eta_s/l_p k_B T\ ln\ (2\xi/d)](2L/3\pi)^4$ where $d$ and $L$ are the filament diameter and length and $\eta_s$ is the solvent viscosity. A representative average actin filament length of $L \approx 5$ µm yields $\tau_b \approx 39$ ms and $\tau_b^{-1} \approx 25$ s$^{-1}$, a rate that is comparable to or faster than all measured strain rates. This relation suggests that actin filaments are able to bend in response to the applied strains to dissipate stress, rendering bending modes to be likely contributors to the composite response.

## Section 3. Mechanical Circuit

To capture the dynamic force-response of the composite network to local deformation via the motion of an optically-trapped particle, we constructed a mechano-equivalent circuit model as follows:

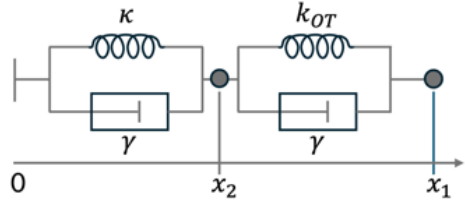

where $x_1$ denotes the position of the center of mass position of the optical trap and $x_2$ denotes the center of mass position of the particle. Within the model, we assume

$$\frac{dx_1}{dt} = v \qquad \text{S12}$$

$$x_1 = vt + L_0 \qquad \text{S13}$$

where $L_0$ is $x_1$ at $t = 0$. Note that this construction assumes a different frame of reference than in the experimental case, in which the optical trap, rather than the stage, is displaced at a fixed speed $v$. Thus, forces within the model are calculated as a function of $x_1$, whereas the experimental data are plotted as force versus stage displacement $x$. We treat these two independent parameters as equivalent. By balancing the forces in the two Kelvin-Voigt modules, we found

$$k_{trap}(x_1 - x_2) + \gamma \frac{d}{dt}(x_1 - x_2) = \kappa(x_2) + \gamma \frac{d}{dt}(x_2) \qquad \text{S14}$$

which simplifies to

$$\frac{dx_2}{dt} + \frac{(\kappa + k_{trap})}{2\gamma} x_2 = \frac{k_{trap} v}{2\gamma} t + \frac{(k_{trap} L_0 + \gamma v)}{2\gamma} \qquad \text{S15}$$

Solving for $x_2$ under the assumption that at $t = 0, x_2 = x_1 = L_0 \sim 0$ we get

$$x_2 = \frac{k_{trap}}{k_{trap} + \kappa} vt + \frac{\gamma v}{k_{trap} + \kappa}\left(1 - \frac{2k_{trap}}{k_{trap} + \kappa}\right)\left(1 - e^{-\left(\frac{k_{trap} + \kappa}{2\gamma}\right)t}\right) \qquad \text{S16}$$

To find the force as a function of trap displacement, $x_1$ we asserted $F(x_1) = k_{trap}(x_1 - x_2)$. Substituting $t = \frac{x_1}{v}$ in Eqn. S16 above, we obtain

$$F(x_1) = \frac{k_{trap} \kappa}{k_{trap} + \kappa} \cdot x_1 - \gamma v k_{trap}\left(\frac{\kappa - k_{trap}}{(k_{trap} + \kappa)^2}\right)\left(1 - e^{-\left(\frac{k_{trap} + \kappa}{2\gamma}\right)x_1}\right)$$

which provides a predictive relationship between stage displacement, speed, and force.

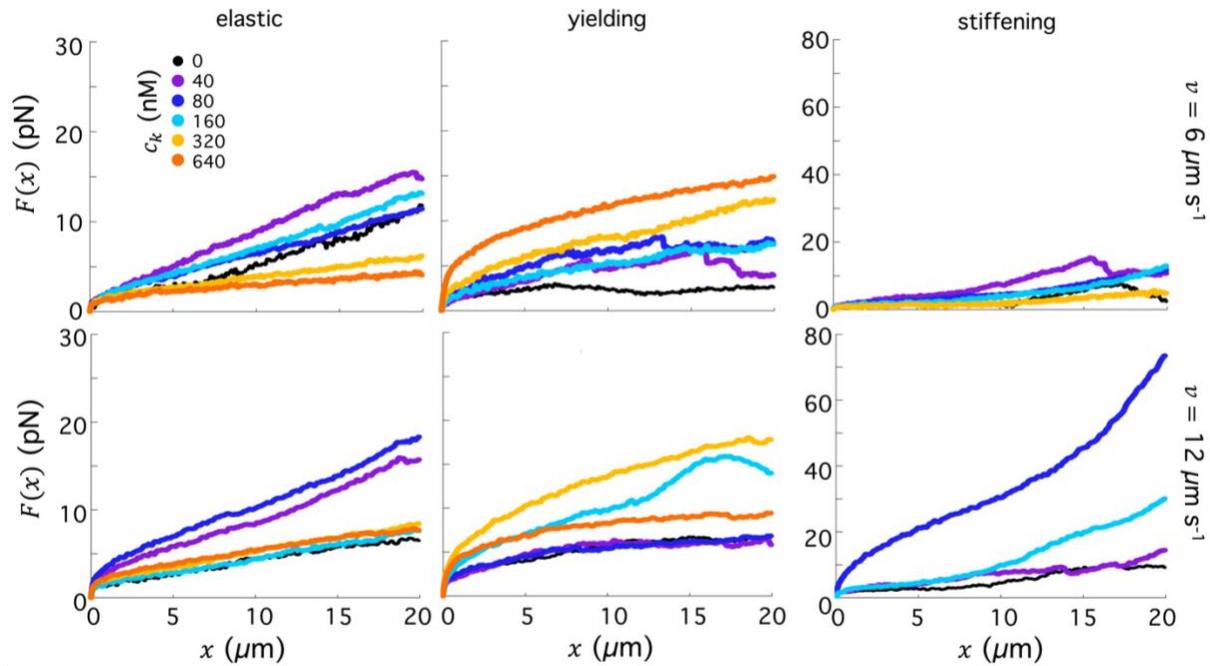

**Figure S2. Average force traces classified by response type for strains of speed $v = 6$ μm s$^{-1}$ (top) and $v = 12$ μm s$^{-1}$ (bottom).** Average force $F(x)$ versus stage position $x$ for each kinesin concentration $c_k$, listed and color-coded according to the legend in the top left panel. Force traces classified as elastic (left), yielding (middle), and stiffening (right) are averaged separately. The highest kinesin concentrations lacked stiffening traces at some speeds, in which case no data are shown in the righthand panel. The data for $v = 24$ μm s$^{-1}$ are shown in Figure 4A.

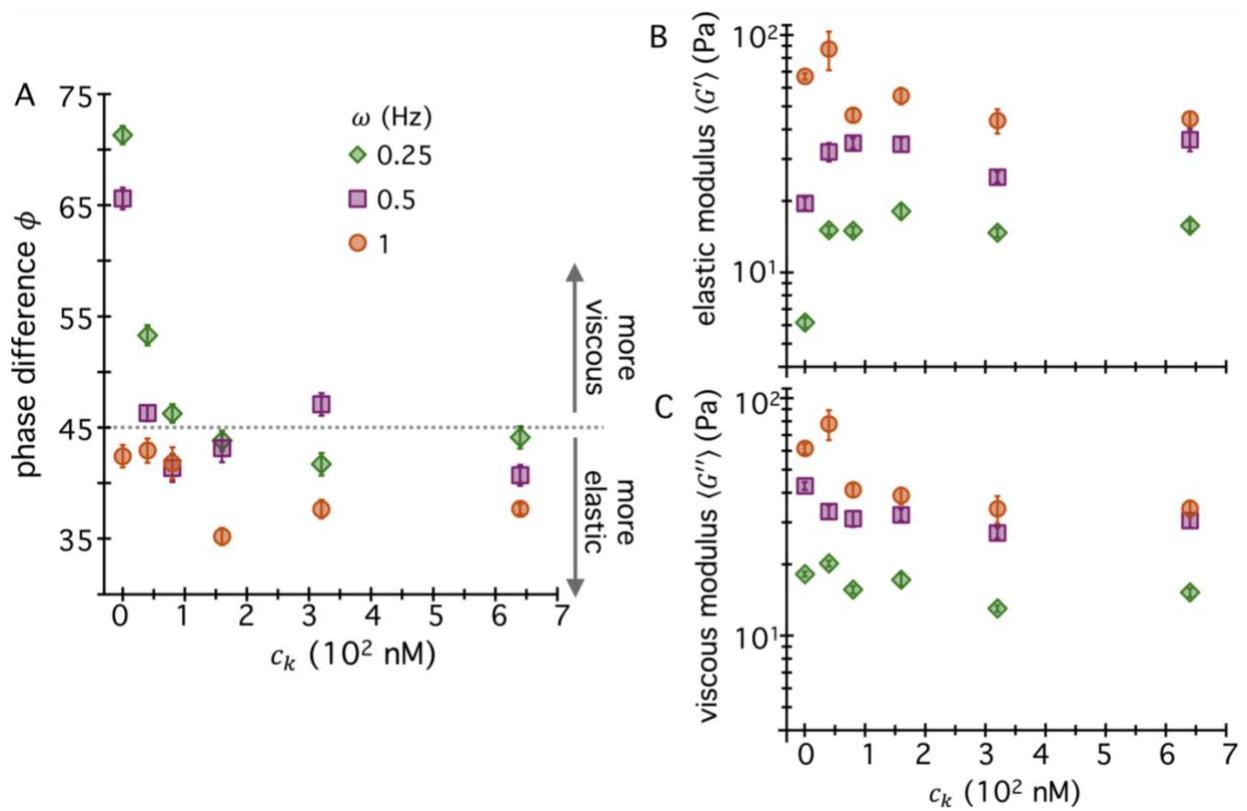

**Figure S3. Viscoelastic moduli determined from simulated oscillatory forcing of a bead through 2D active cytoskeletal composites at varying oscillation frequencies.** (A) The phase difference $\phi$ (in units of degrees) between the force oscillation and bead displacement oscillation as a function of kinesin concentration $c_k$. Force oscillations were performed at frequencies of $\omega =$ 0.25 (green diamonds), 0.5 (purple squares) and 1 (red circles) Hz. $\phi = 0$ and $\phi = 90$ correspond to purely elastic and viscous behaviour, respectively, as indicated by the arrows. The dashed line at $\phi = 45°$ indicates equal viscous and elastic contributions. (B,C) The elastic and viscous moduli, $G'$ (B) and $G''$ (C), computed from the phase difference, force $F$, and bead displacement $x$ as described in SI Section 1.

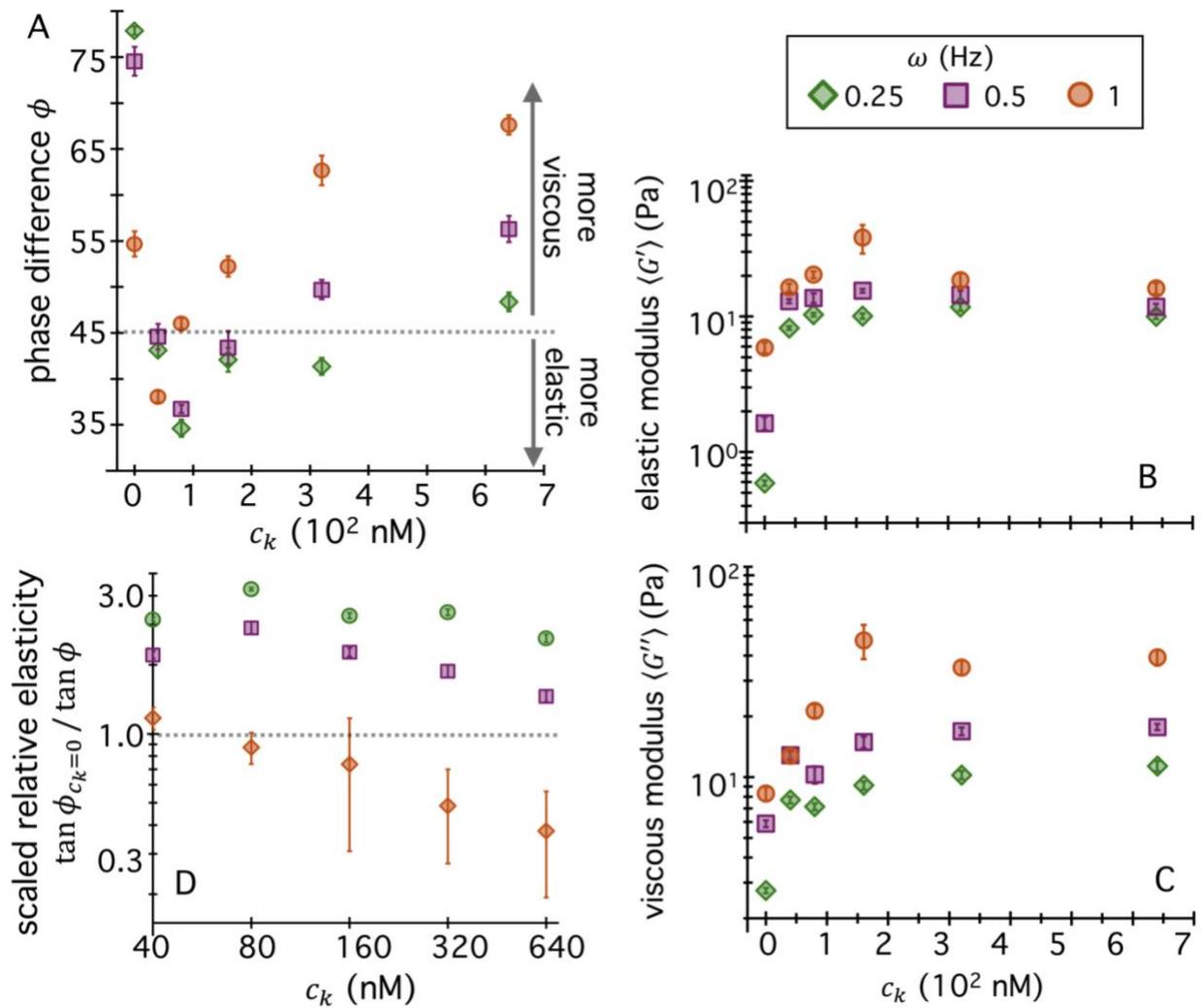

**Figure S4. Viscoelastic moduli determined from simulated oscillatory forcing of a bead through active microtubule networks (no actin) at varying oscillation frequencies.** Same simulations and data as presented in Figure 4C-E and S4, but without actin. All other concentrations and parameters are the same. (A) The phase difference $\phi$ (in units of degrees) between the force oscillation and bead displacement oscillation as a function of kinesin concentration $c_k$. Force oscillations were performed at frequencies of $\omega = 0.25$ (green diamonds), 0.5 (purple squares) and 1 (red circles) Hz. $\phi = 0$ and $\phi = 90$ correspond to purely elastic and viscous behavior, respectively, as indicated by the arrows. The dashed line at $\phi = 45°$ indicates equal viscous and elastic contributions. (B,C) The elastic and viscous moduli, $G'$ (B) and $G''$ (C), computed from the phase difference, force $F$, and bead displacement $x$ as described in SI Section 1. (D) Scaled relative elasticity, computed as the inverse loss tangent $[\tan\phi]^{-1}$ normalized by the corresponding $c_k = 0$ value, indicated by the dashed horizontal line, versus kinesin concentration. Color and symbol scheme is the same for all panels and shown above panel B. Error bars in all panels correspond to standard deviation of bootstrapped ensembles.

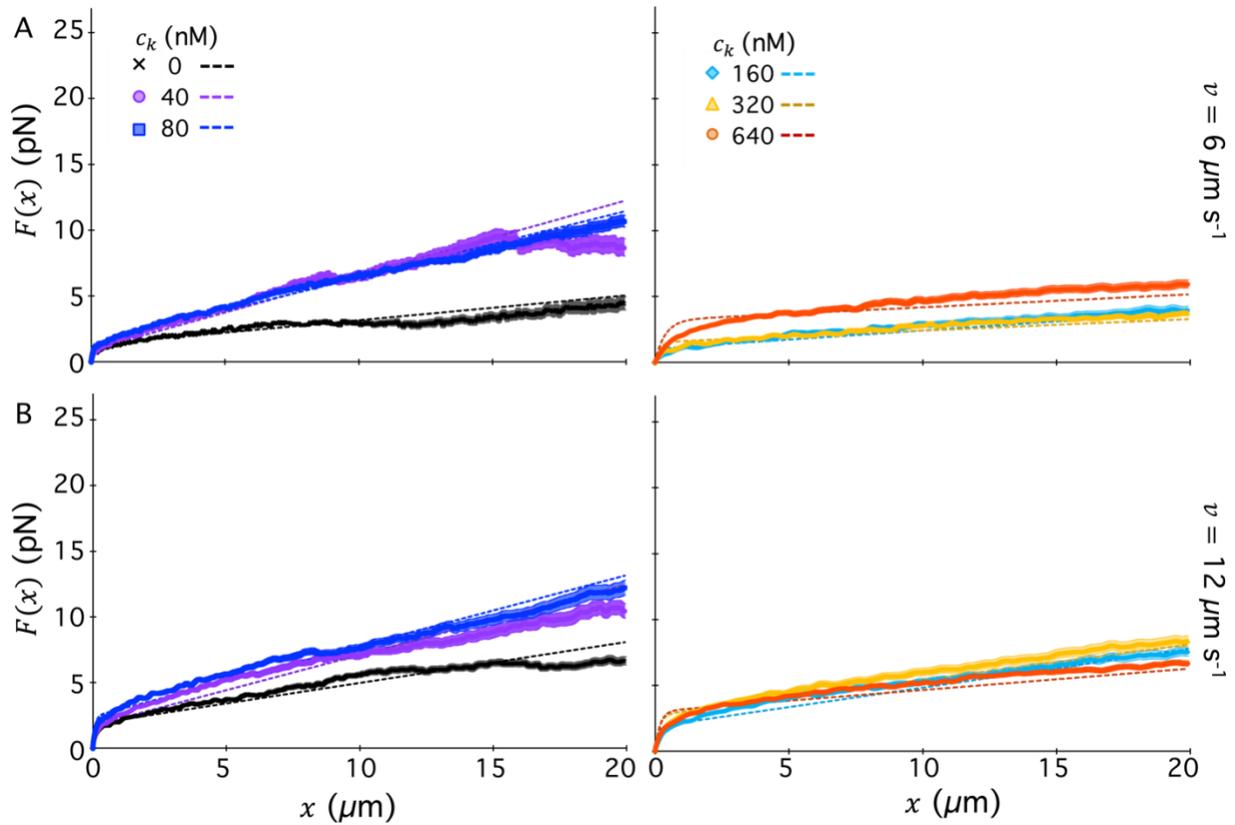

**Figure S5. Mechanical circuit model captures the viscoelastic behavior of elastic and yielding response classes.** Force $F(x)$ versus stage position $x$, averaged across all elastic and yielding traces for each kinesin concentration $c_k$, listed and color-coded according to the legend, for (A) $v = 6$ μm s$^{-1}$ and (B) $v = 12$ μm s$^{-1}$. Error bars denote standard error of the mean. Dashed lines are fits to the equation of motion for the mechanical circuit depicted in Fig 5A. Data for $v = 24$ μm s$^{-1}$ is shown in Fig 5B.

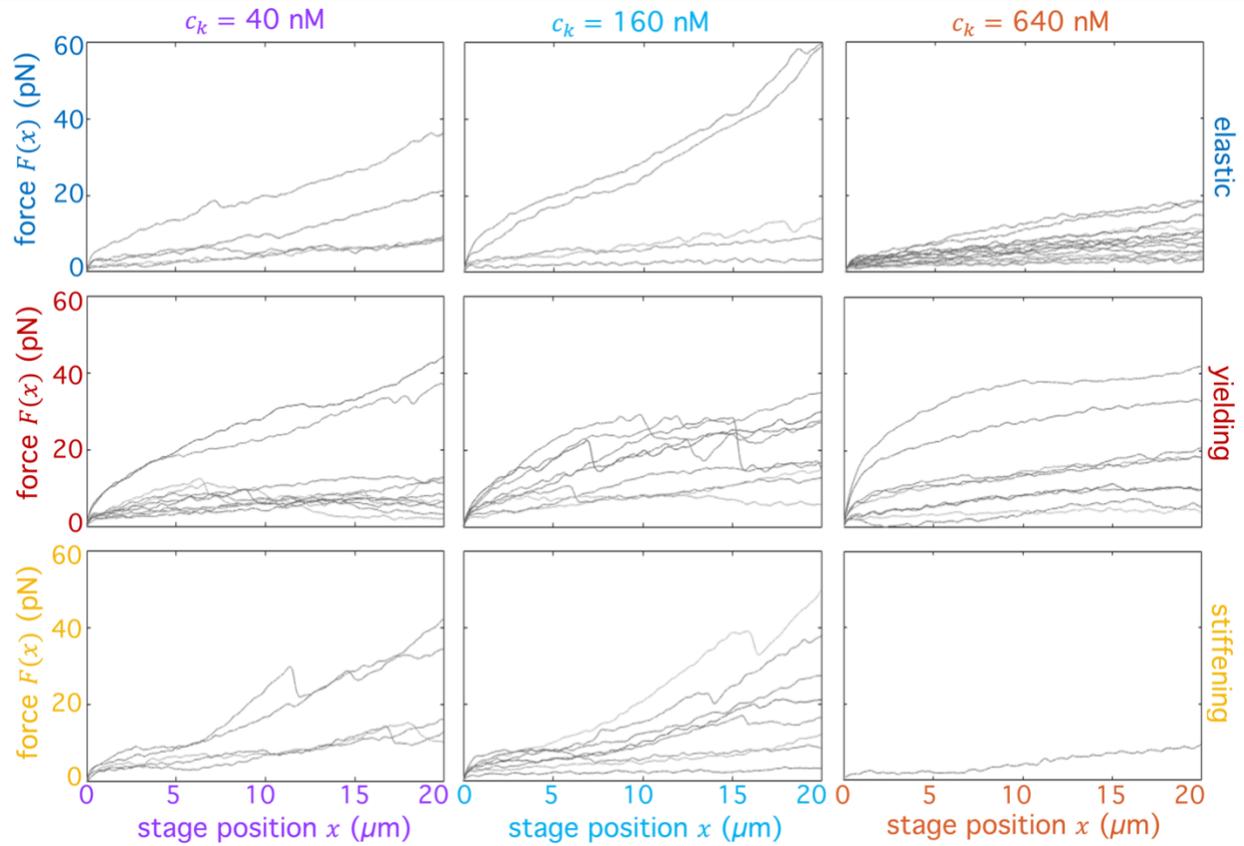

**Figure S6. Ensemble of force traces comprising averages shown in Figure 4A.** Individual force traces measured in response to strains with $v = 24$ μm s$^{-1}$ in composites with kinesin concentrations of $c_k = 40$ nM (left, purple labels), 160 nM (middle, cyan labels), 640 nM (right, orange labels). Traces shown compromise the average curves shown in Figure 4A for elastic (top, blue labels), yielding (middle, red), and stiffening (bottom, gold) responses at the given kinesin concentrations.